\documentclass[12pt]{article}

\usepackage[utf8]{inputenc} 
\usepackage[T1]{fontenc}    
\usepackage{url}            
\usepackage{booktabs}       
\usepackage{amsfonts}       
\usepackage{nicefrac}       
\usepackage{microtype}      
\usepackage{lipsum}
\usepackage{fancyhdr}       
\usepackage{graphicx}       
\graphicspath{{media/}}     

\usepackage{natbib}
\usepackage{enumitem}
\setlist[itemize]{topsep=0pt, partopsep=0pt, itemsep=0pt, parsep=0pt}

\usepackage{comment}
\usepackage{amsthm}
\usepackage{amsmath}
\usepackage{amssymb}
\usepackage{algorithm}
\usepackage{algorithmic}
\usepackage{multirow}
\usepackage{array}
\usepackage{babel}
\usepackage{setspace}
\usepackage{soul}
\usepackage{pdflscape}
\usepackage{tabularx}

\usepackage{xcolor}
\sethlcolor{yellow}
\usepackage[usestackEOL]{stackengine}
\usepackage[left=1in, right=1in, top=1in, bottom=1in]{geometry}
\usepackage{float}

\newtheorem{theorem}{Theorem}
\newtheorem{assumption}{Assumption}
\usepackage{makecell}
\newcolumntype{o}{>{\centering\arraybackslash\hsize=0.12\hsize}X}
\newcolumntype{s}{>{\centering\arraybackslash\hsize=0.2\hsize}X}
\newcolumntype{m}{>{\centering\arraybackslash\hsize=0.3\hsize}X}
\newcolumntype{b}{>{\centering\arraybackslash\hsize=1\hsize}X}
\newcolumntype{B}{>{\centering\arraybackslash\hsize=1.3\hsize}X}

\newcolumntype{x}{>{\arraybackslash\hsize=0.2\hsize}X}
\newcolumntype{y}{>{\arraybackslash\hsize=1\hsize}X}
\newcolumntype{z}{>{\arraybackslash\hsize=1.3\hsize}X}

\newcolumntype{M}[1]{>{\centering\arraybackslash}m{#1}}

\usepackage[colorlinks=true, allcolors=blue]{hyperref}

\newcommand{\blind}{1}

\begin{document}

\def\spacingset#1{\renewcommand{\baselinestretch}%
{#1}\small\normalsize} \spacingset{1}


\if1\blind
{
  \title{\bf 
  A new statistical approach for joint modeling of longitudinal outcomes measured in electronic health records with clinically informative presence and observation processes}
  \vspace{-5cm}
  \author{  
  Jiacong Du\\
  Department of Biostatistics\\
  University of Michigan, USA\\
  \and
  Xu Shi\\
  Department of Biostatistics\\
  University of Michigan, USA\\
  \and \vspace{-1.25cm} \\    
  Bhramar Mukherjee\thanks{Corresponding author} \\
  Department of Biostatistics\\
  Department of Chronic Disease Epidemiology\\
  Department of Statistics and Data Science\\
  Yale University, USA\\ 
  \vspace{-0.8 cm}
    }
  \maketitle
} \fi

\if0\blind
{
  \bigskip
  \bigskip
  \bigskip
  \begin{center}
    {\LARGE\bf A new statistical approach for joint modeling of longitudinal outcomes measured in electronic health records with clinically informative presence and observation processes}
\end{center}
  \medskip
} \fi

\vspace{-1.75em} 
\begin{abstract}
    Biobanks with genetics-linked electronic health records (EHR) have opened up opportunities to study associations between genetic, social, or environmental factors and longitudinal lab biomarkers. However, in EHRs, the timing of patient visits and the recording of lab tests often depend on patient health status, referred to as informative presence (IP) and informative observation (IO), which can bias exposure–biomarker associations. Two gaps remain in EHR-based research: (1) the performance of existing IP-aware methods is unclear in real-world EHR settings, and (2) no existing methods handle IP and IO simultaneously. To address these challenges, we first conduct extensive simulation studies tailored to EHR-specific IP patterns to assess existing methods. We then propose a joint modeling framework, EHRJoint, that simultaneously models the visiting, observation, and longitudinal biomarker processes to address both IP and IO. We develop a computationally efficient estimation procedure based on estimating equations and provide asymptotically valid inference. Simulations show that EHRJoint yields unbiased exposure effect estimates under both IP and IO, while existing methods fail. We apply EHRJoint to the Michigan Genomics Initiative data to examine associations between repeated glucose measurements and two exposures: genetic variants and educational disadvantage.
\end{abstract}
\noindent%


\newpage

\spacingset{1.45} 

\section{Introduction}

Electronic health records (EHR) offer a valuable resource for studying exposure-biomarker associations using routinely collected lab measurements. Large-scale biobanks, such as the UK Biobank \citep{bycroft2017genome} and All of Us \citep{all2019all}, further enhance this research by linking genetic data with EHR, enabling exploration of exposure-biomarker association, where the exposure could be genetic, environmental or socioeconomic. The agnostic search for associations between lab biomarkers and risk factors across all available lab markers in a biobank is known as laboratory biomarker-wide association studies (LabWAS) \citep{goldstein2020labwas,dennis2021clinical}. In LabWAS, longitudinal biomarker measurements are typically summarized using mean, median, minimum, or maximum across repeated measurements. These summary statistics are then used as outcomes in regression models. While this approach has led to important discoveries, it overlooks within-patient variation, and could be highly sensitive to the choice of summary statistic and number of measurements. Beyond LabWAS, in associating a longitudinal biomarker with exposures (both baseline and time varying), not using the full longitudinal data and reducing it to a single summary outcome is inferentially sub-optimal.

While there are statistical methods designed for longitudinal data analysis \citep{verbeke1997linear,diggle2002analysis}, unique challenges exist for EHR data. Patient visits are not pre-scheduled but occur opportunistically based on patients’ underlying health conditions and access to healthcare \citep{goldstein2016controllingEHRencounter}. Furthermore, biomarkers of interest may not be recorded at every visit, leading to informative missing data patterns \citep{wells2013strategies, haneuse2016general}. These challenges arise from two interrelated processes: the \textit{visiting process} and the \textit{observation process}. The visiting process determines when a patient comes to the clinic, while the observation process determines whether the biomarker of interest is recorded during a visit. 
These concepts have been used in the literature to characterize the unique features of EHR data
\citep{goldstein2016controllingEHRencounter, goldstein2019and, sisk2021informative, harton2022informative}. 
\newpage 
\noindent In many cases, both processes may depend on measured (e.g., patient characteristics), unmeasured variables, or the biomarker itself. When the visiting process depends on the biomarker—either directly or indirectly through measured variables and unmeasured factors, we refer to this phenomenon as \textit{informative presence} (IP) of the patient visit. Similarly, the observation process may also depend on the biomarker in the same way. We refer to this as \textit{informative observation} (IO) related to the biomarker.

\begin{figure}[h]
    \centering
    \includegraphics[width=0.45\linewidth]{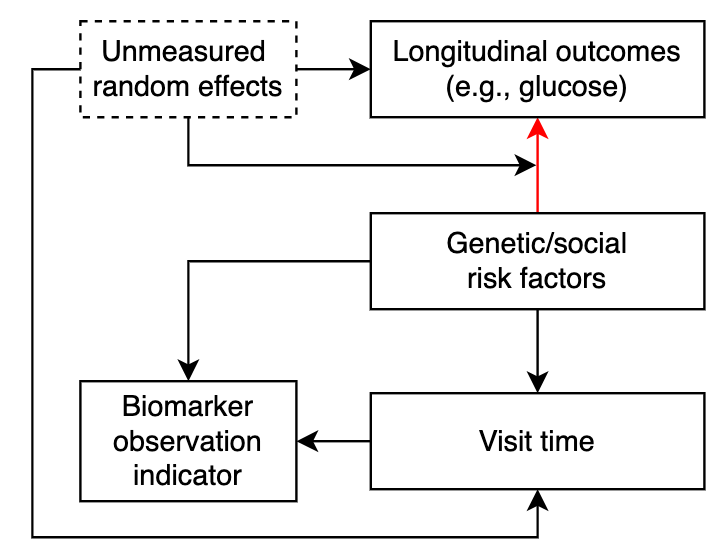}
    \caption{Directed Acyclic Graph illustrating the potential bias arising from conditioning on observed data in longitudinal EHR analysis. Random effects are unmeasured and presented in the dotted box. Conditioning on the observed data opens the pathway from the exposure to the outcome.}
    \label{fig:DAG}
\end{figure}

Ignoring IP and IO can result in biased exposure effect estimates \citep{goldstein2016controllingEHRencounter,mcculloch2016biased,neuhaus2018analysis}. To illustrate this, \textbf{Figure \ref{fig:DAG}} presents a conceptual representation through Directed Acyclic Graph (DAG) of a specific case of IP and IO. We focus on the association between an exposure (could be genetic or environmental) and a longitudinal biomarker outcome (EHR-measured glucose levels). We include random effects to account for individual-level variation and heterogeneity in the exposure effect across individuals. These random effects may also capture variability in patient clinic visit frequency. Additionally, whether the biomarker is observed acts as a collider in the exposure-outcome relationship: the exposure affects biomarker observation both directly and indirectly through visit time, while the biomarker outcome impacts its observation via unmeasured random effects and visit time. Consequently, conditioning on this collider opens a pathway between the exposure and outcome, and thus biases the estimated effect.

With the growing adoption of EHR systems, several methods have been proposed to modify standard mixed-effects model analysis \citep{lin2001semiparametric,liang2009joint,burvzkova2007longitudinal,buuvzkova2010longitudinal,gasparini2020mixed}. While these proposals account for IP, they often assume that the biomarker of interest is observed at every visit, or limit the analysis to only those visits where the biomarker is recorded. As a result, they exclude data from patients who have clinical visits but without biomarker measurements, even though such data are abundant in the EHR. 
Therefore, handling missing biomarker measurements is particularly important, especially when the missingness arises from IO. Furthermore, many existing IP-aware methods are originally developed for classical longitudinal studies with fewer subjects, shorter follow-up periods, and fewer observations per subject, while EHR data often span multiple decades and contain hundreds of visits and records per subject. Although some methods have been reviewed and compared  \citep{tan2014regression,neuhaus2018analysis}, a comprehensive evaluation of a large set of methods within the EHR context is still missing. 

The goals of this paper are twofold. First, we conduct extensive simulation studies tailored under the EHR-specific IP patterns to evaluate the performance of existing methods. These include the summary statistics approach commonly used in LabWAS, standard mixed-effects models, and methods that address IP via inverse intensity weighting or joint modeling of the longitudinal and visiting processes.
Second, recognizing the lack of methods that address both IP and IO, we propose a novel tripartite joint model with three interconnected sub-models: one for the visiting process to model patient visit times, one for the observation process to capture whether the biomarker of interest is recorded at each visit, and one for the longitudinal process to describe the biomarker trajectory over time.
Building on \cite{liang2009joint}'s framework, our proposed joint model effectively captures correlations between the visiting and longitudinal processes resulting from unmeasured variables. To handle IO, we incorporate a third model component that accounts for whether the biomarker is observed at each visit. Through this embedded structure, our joint model handles a hierarchical missingness mechanism for the longitudinal biomarker outcome. The first level of missingness 
\newpage 
\noindent occurs when patients do not visit the clinic, while the second level arises if a lab test is not conducted during a recorded visit. In contrast, existing methods \citep{lin2001semiparametric,liang2009joint,burvzkova2007longitudinal,buuvzkova2010longitudinal,gasparini2020mixed} typically assume that biomarker measurements are always observed during visits. As a result, these methods account only for the first level of missingness. We also introduce a computationally efficient estimation procedure using step-wise estimating equations for the tripartite joint model and derive the asymptotic distribution of the resulting estimators.

Our motivating dataset is from the longitudinal biorepository known as the Michigan Genomics Initiative (MGI). Participants in the MGI cohort were primarily recruited through the Michigan Medicine health system and provided consent to link their EHRs and genetic data for research. Recruitment started in 2012, and MGI currently has over 100,000 consented participants. The EHR data from MGI include patient diagnoses, demographics, lifestyle and behavioral risk factors, and lab measurements. Our objective is to study the association between exposures (e.g., genetic variants, social risk factors) and longitudinal lab measurements (e.g., blood glucose levels). The study period is defined as the first five years since enrollment. After applying the selection and exclusion criteria, the total analytical sample consists of 26,442 patients. Among these, 90.5\% of patients have at least one recorded visit, while only 40.8\% have at least one glucose measurement. The number of visits per patient varies widely, ranging from 0 to 190 visits, with a median of 8 visits per patient. In contrast, glucose measurements were less frequent. Among the 40.8\% of patients with glucose measurements, the median number of measurements is 3 (min: 1, max:162).

This paper is organized as follows: 
Section 2 presents a mathematical framework that characterizes how biomarker measurements are collected in the EHR. Within this framework, the visiting process, observation process and longitudinal process are clearly defined. Next, we review existing methods and highlight one in particular for handling only IP. Next, we propose a method that handles both IP and IO, and describe the estimation procedure and 
\newpage 
\noindent its theoretical properties. In Section 3, we perform three sets of simulation studies. The first examines performance under IP-only scenarios using existing methods. The second introduces both IP and IO, and compare our proposed method with existing methods. The third set studies the robustness of our proposed method to model misspecification in scenarios involving both IP and IO.
In Section 4, we investigate the associations between repeated glucose measurements and two exposures: genetic variants and educational disadvantage. 
We conclude with a summary of our findings and recommendations in Section 5. 

\section{Methods}\label{section_methods}

\subsection{Framework} \label{section_methods_framework}

\begin{figure}[h]
    \centering
    \includegraphics[width=1\linewidth]{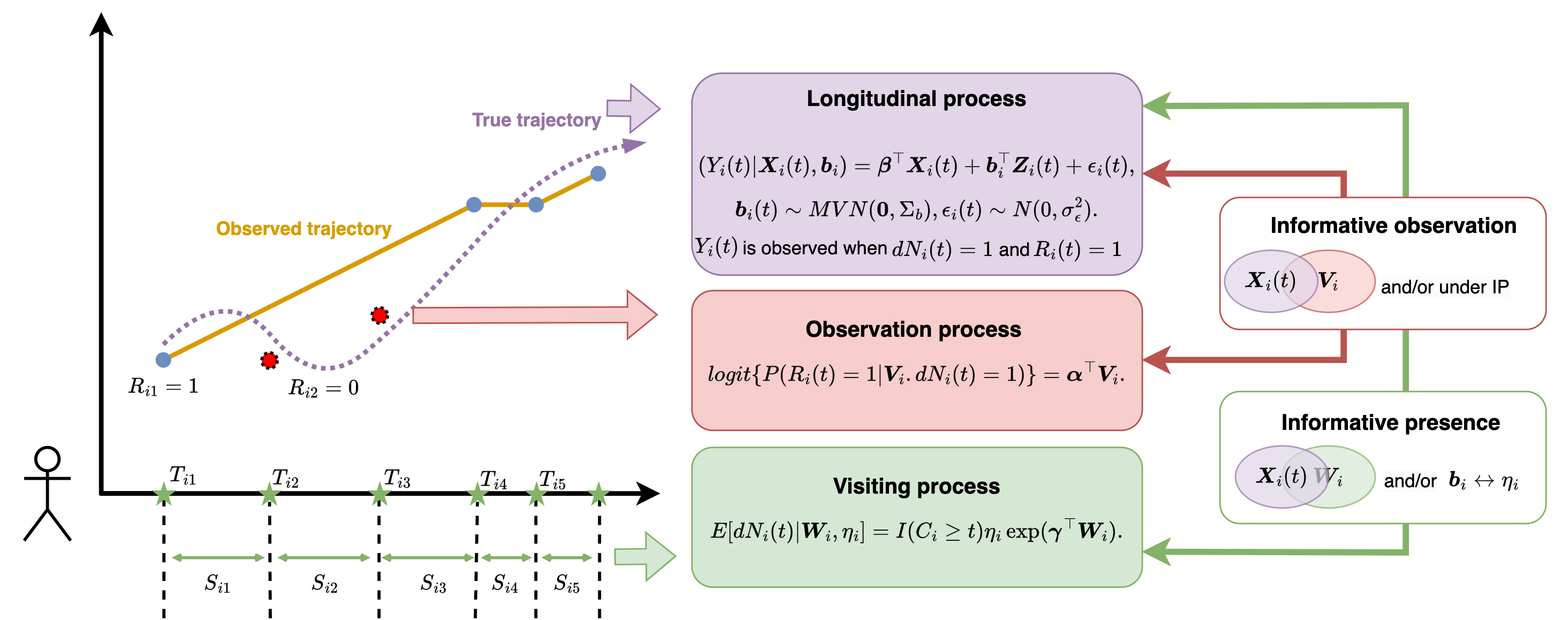}
    \caption{A framework of characterizing the collection of longitudinal biomarker in the EHR and when informative presence (IP) and informative observation (IO) occur.}
    \label{fig_framework}
\end{figure}

Let $n$ denote the number of patients. The collection of biomarker measurements in the EHR system over time can be characterized by three main processes: the visiting process, the observation process, and the longitudinal process, as illustrated in \textbf{Figure \ref{fig_framework}}. 

The \textbf{visiting process} captures the temporal pattern of patients' clinical visits. Let $T_{ij}$ denote the timestamp for the $j$-th visit of the $i$-th patient, where $j=1,...,n_i$ and $i=1,..., n$. Using counting process notation, let $N_i(t) = \sum_{j=1}^{n_i}I(T_{ij}\leq t)$ be the number of visits up to time $t$ for the $i$-th patient. We assume the following frailty model for the visiting process:
\begin{equation*}\label{framework_VPmodel}
    \mathbb{E}[dN_i(t)|\boldsymbol{W}_i,\eta_i] = \xi_i(t) \eta_i \exp
    \big(
    \boldsymbol{\gamma}^\top \boldsymbol{W}_i
    \big)d\Lambda_0(t),
\end{equation*}
where $dN_i(t)$ is the visit indicator, which is $1$ if the visit occurs at $t$ and $0$ otherwise. $\xi_i(t)=I(t\leq C_i)$ is the censoring indicator with $C_i$ being the censoring time. $\Lambda_0(t)$ is the baseline intensity function. 
$\boldsymbol{W}_i$ is a set of patient-level covariates that affect their clinical visit intensity. In practice, there might be time-varying covariates that impact the visiting process. However, for the purpose of our current paper, we assume that $\boldsymbol{W}_i$ only contain time-invariant variables, which is sufficient to demonstrate the issues of IP and IO. The frailty term $\eta_i$ is an unobserved latent variable that quantifies how visit intensity for an individual deviates from the population average visit intensity.

The \textbf{observation process} determines whether the biomarker is observed or not at a given visit time. Let $R_i(t)$ be the recording indicator of the biomarker, which takes value $1$ if the biomarker is recorded at time $t$ and $0$ otherwise. Since the biomarker is only recorded when patients visit the clinic, it is natural to model $R_i(t)$ conditional on the visiting process. Therefore, we assume the following logistic regression model for $R_i(t)$:
\begin{equation*}\label{framework_OPmodel}
    logit \{\mathbb{P}(R_i(t)=1|dN_i(t)=1,\boldsymbol{V}_i)\} = \boldsymbol{\alpha^\top \boldsymbol{V}_i}.
\end{equation*}
To be computationally efficient while effectively illustrating the main issues with modeling of IP and IO, we adopt a population-averaged marginal model, assuming $\boldsymbol{V}_i$ consists of time-invariant covariates that affect the probability of recording the biomarker. Since existing methods assume that $R_i(t) = 1$ when $dN_i(t) = 1$, directly applying these methods that ignore IO means that patient data with $R_i(t) = 0$ are excluded, even though such data are abundant in the EHR.

The \textbf{longitudinal process} characterizes the biomarker trajectory over time. Let $Y_i(t)$ denote the biomarker value for the $i$-th patient at time $t$. 
Note that we do not observe the biomarker $Y_i(t)$ 
for any $t$, but only when patient $i$ visits the clinic and when $Y$ is recorded. For example, consider glucose levels as the biomarker $Y$. If a patient does not visit the clinic at $t$, i.e., $dN_i(t)=0$, their glucose levels cannot be observed. Furthermore, even if they visit the clinic, their glucose levels may still remain unrecorded if a lab test is not ordered by clinicians or conducted, i.e., $R_i(t)=0$. 
Although the observed trajectory of the longitudinal biomarker is discrete, it is often assumed that the true underlying trajectory is continuous in time. With continuous outcomes, we assume a linear mixed-effects model for the longitudinal biomarker process:
\begin{equation*}\label{framework_Ymodel}
    (Y_i(t) | \boldsymbol{X}_i(t), \boldsymbol{b}_i)= 
    \boldsymbol{\beta}^\top \boldsymbol{X}_i(t) + \boldsymbol{b}_i^\top \boldsymbol{Z}_i(t) + \epsilon_i(t),
\end{equation*}
\noindent where $(\cdot|\cdot)$ denotes conditional distribution, $\epsilon_i(t) \sim \mathcal{N}(0,\sigma^2)$, $\boldsymbol{X}_i(t)$ is a vector of patient-level covariates at time $t$, which could be time-varying or time-invariant, and $\boldsymbol{Z}_i(t)$ is a subset of $\boldsymbol{X}_i(t)$. $\boldsymbol{\beta}$ is the fixed effect of $\boldsymbol{X}_i(t)$, and $\boldsymbol{b}_i \sim MVN (\boldsymbol{0},\Sigma_b)$ is a $q$-dimensional random effect vector of $\boldsymbol{Z}_i(t)$. 

With the three defined processes and their associated models, we can conceptualize how IP and IO arise from a modeling perspective. In \textbf{Figure \ref{fig_framework}}, IP occurs when the visiting and longitudinal processes are dependent, through observed covariates $\boldsymbol{W}_i$ and $\boldsymbol{X}_i(t)$, or/and through unmeasured random effects $\eta_i$ and $\boldsymbol{b}_i$. IO occurs when the observation and longitudinal processes are dependent, through measured covariates $\boldsymbol{V}_i$ and $\boldsymbol{X}_i(t)$, or/and when the coefficient $\boldsymbol{\alpha}$ is non-zero under IP. 
In addition, the true underlying processes could be complicated and the outlined parametric stochastic models may not fully capture these dynamics. Such scenarios are explored later through simulation studies.

\subsection{Existing approaches}\label{section_existing_methods}

\noindent \textbf{VA-LME.} A straightforward approach to analyzing longitudinal EHR data is to fit a visit-adjusted linear mixed-effects (VA-LME) model that includes the number of visits up to time $t$ as a time varying covariate:
\begin{equation*}
    (Y_i(t) | \boldsymbol{X}_i(t), N_i(t), \boldsymbol{b}_i)= 
    \boldsymbol{\beta}^\top \boldsymbol{X}_i(t) + \beta_n N_i(t) + \boldsymbol{b}_i^\top \boldsymbol{Z}_i(t) + \epsilon_i(t),
\end{equation*}
\noindent where $\boldsymbol{b}_i \sim \mathcal{N}(0,\Sigma_b)$, $\epsilon_i(t) \sim \mathcal{N}(0,\sigma^2)$ and $N_i(t)$ is the number of visits for patient $i$ before time $t$. As noted by \citet{goldstein2016controllingEHRencounter}, $N_i(t)$ serves as a proxy for illness severity and helps mitigate bias due to IP.

\noindent \textbf{JMVL-LY.} Traditional linear mixed-effects models assume non-stochastic observation times, which is often violated in EHR settings due to irregular visit patterns. To address this, joint modeling of the visiting and longitudinal processes (JMVL) has been proposed. Specifically, \cite{lin2001semiparametric} introduces a proportional rate model for the visiting process and a marginal semi-parametric model for the longitudinal biomarker (JMVL-LY):
\begin{align*}
    \mathbb{E}[dN_i(t)|\boldsymbol{W}_i(t)] & = \xi_i(t) \exp
    \big(
    \boldsymbol{\gamma}^\top \boldsymbol{W}_i(t)
    \big)d\Lambda_0(t), \\
    \mathbb{E}[Y_i(t)|\boldsymbol{X}_i(t)] & = 
    \beta_0(t) + \boldsymbol{\beta}^\top \boldsymbol{X}_i(t),
\end{align*}
\noindent where $\beta_0(t)$ is a function of time, and $\boldsymbol{W}_i(t) \subseteq \boldsymbol{X}_i(t)$ represents time-varying covariates. JMVL-LY assumes that $N_i(t)$ and $Y_i(t)$ are independent conditional on $\boldsymbol{W}_i(t)$, so that the visiting and longitudinal processes are linked only through measured variables $\boldsymbol{W}_i(t)$. Details about parameter estimation under this model are provided in Supplementary Section S1.

\noindent \textbf{IIRR-weighting.} An alternative approach to address IP is through inverse-intensity-rate-ratio (IIRR) weighting \citep{burvzkova2007longitudinal}. This method uses a proportional rate model for the visiting process and a parametric marginal model for the longitudinal process:
\begin{align*}
    \mathbb{E}[dN_i(t)|\boldsymbol{W}_i(t)] &= \xi_i(t) \exp
    \big(
    \boldsymbol{\gamma}^\top \boldsymbol{W}_i(t)
    \big)d\Lambda_0(t),\\
    \mathbb{E}[Y_i(t)|\boldsymbol{X}_i(t)] &= 
    \boldsymbol{\beta}^\top \boldsymbol{X}_i(t).
\end{align*}
\noindent Unlike JMVL-LY, where $\boldsymbol{W}_i(t)$ is restricted to a subset of $\boldsymbol{X}_i(t)$, the IIRR-weighting method allows $\boldsymbol{W}_i(t)$ to include the history of covariates, auxiliary factors, and past biomarker information. The estimation procedure weights each longitudinal observation inversely by the relative intensity of being observed, i.e., $\exp \big( \boldsymbol{\gamma}^\top \boldsymbol{W}_i(t) \big)$.

\noindent \textbf{JMVL-G.} 
Satisfying the conditional independence assumption in JMVL-LY is challenging in the EHR, since not all factors leading to clinical visits can be measured. Therefore, JMVL-G relaxes this assumption by incorporating a shared random effect that links the visiting and longitudinal processes \citep{gasparini2020mixed}. JMVL-G uses a proportional hazard model for the inter-visit time $S_{ij}$, where $S_{ij} = T_{ij}-T_{i(j-1)}$. The joint model is specified as:
\begin{align*}
    r(s_{ij}|\boldsymbol{W}_{ij},b_{1i}) &= r_0(s_{ij})\exp(\boldsymbol{\gamma}^\top \boldsymbol{W}_{ij}+b_{1i}),\\
    (Y_{ij}|\boldsymbol{X}_{ij},b_{1i},b_{2i}) &= 
    \boldsymbol{\beta}^\top \boldsymbol{X}_{ij}+\beta_b b_{1i} + b_{2i} +\epsilon_{ij},
\end{align*}
\noindent where $\epsilon_{ij} \sim \mathcal{N}(0,\sigma^2)$, $(b_{1i},b_{2i})^\top \sim \mathcal{N}(\boldsymbol{0},\Sigma_b)$, and $r_0(s_{ij})$ is the baseline intensity function. In addition to measured variables $\boldsymbol{W}_{ij}$ and $\boldsymbol{X}_{ij}$, the visiting and longitudinal processes are also linked through the unmeasured random effect $b_{1i}$.

\noindent \textbf{JMVL-Liang.} Unlike JMVL-G, which links the visiting and longitudinal processes through a single shared random effect, the joint model proposed by \cite{liang2009joint} allows all random effects in the longitudinal model to depend on a frailty variable in the visiting process model. This approach, referred to as JMVL-Liang, is specified as:
\begin{align*}
    \mathbb{E}[dN_i(t)|\boldsymbol{W}_i,\eta_i] &= \xi_i(t) \eta_i \exp
    \big(
    \boldsymbol{\gamma}^\top \boldsymbol{W}_i
    \big)d\Lambda_0(t),\\
    \mathbb{E}[Y_i(t)|\boldsymbol{X}_i(t),\boldsymbol{b}_i] &= 
    \beta_0(t) + \boldsymbol{\beta}^\top \boldsymbol{X}_i(t) + \boldsymbol{b}_i^\top \boldsymbol{Z}_i(t),
\end{align*}
\noindent where $\boldsymbol{W}_i$ is a vector of time-invariant variables and $\boldsymbol{X}_i(t)$ is a vector of time-varying and time-invariatnt variables. The unmeasured frailty term $\eta_i$ in the visiting process model flexibly accounts for the impact from unmeasured variables on the visit intensity. This feature is useful in practice, where not all factors influencing a patient's decision to visit a clinic are known. 
Additionally, JMVL-Liang assumes $\eta_i$ follows a gamma distribution with mean $1$ and variance $\sigma_{\eta}^2$, and $\mathbb{E}(\boldsymbol{b}_i|\eta_i)=\boldsymbol{\theta}(\eta_i-1)$. Thus, besides measured variables $\boldsymbol{W}_i$ and $\boldsymbol{X}_i(t)$, the two models for $N_i(t)$ and $Y_i(t)$ additionally depend on each other through the linkage between $\eta_i$ and $\boldsymbol{b}_i$. 

\cite{liang2009joint} employs a two-step estimation procedure to separately estimate $\boldsymbol{\gamma}$ and $\boldsymbol{\beta}$. The first step estimates $\boldsymbol{\gamma}$ using the following estimating equation:
\begin{equation} \label{EE_gamma_w}
U(\boldsymbol{\gamma}) = \sum_{i=1}^{n} \int_{0}^{\infty} \left\{ \boldsymbol{W}_i - \overline{\boldsymbol{W}} \right\} dN_i(t)=0,
\end{equation}
\noindent where
$\overline{\boldsymbol{W}} = \frac{\sum_{i=1}^{n} \xi_i(t) \exp\{ \boldsymbol{\gamma}^\top \boldsymbol{W}_i \}\boldsymbol{W}_i}{\sum_{i=1}^{n} \xi_i(t) \exp\{ \boldsymbol{\gamma}^\top \boldsymbol{W}_i \}}$.
In the second step, $\boldsymbol{\beta}$ and $\boldsymbol{\theta}$ are estimated by solving
\begin{equation*}
U(\boldsymbol{\beta},\boldsymbol{\theta}) = \sum_{i=1}^{n} \int_0^\infty 
\begin{pmatrix}
    \boldsymbol{X}_i(t) - \overline{\boldsymbol{X}}(t) \\
    \hat{\boldsymbol{B}}_i(t) - \overline{\hat{\boldsymbol{B}}}(t)
\end{pmatrix} \left\{ Y_i(t) - \boldsymbol{\beta}^\top \boldsymbol{X}_i(t) - \boldsymbol{\theta}^\top  \hat{\boldsymbol{B}}_i(t) \right\} dN_i(t) = 0 ,
\end{equation*}
where $\boldsymbol{B}_i(t) = \boldsymbol{Z}_i(t)  \mathbb{E}[(\eta_i - 1) | C_i, n_i]$ is the random effect after taking the conditional expectation of $\boldsymbol{b}_i|\eta_i,C_i, n_i$. The centering terms $\overline{\boldsymbol{X}}(t) = \frac{\sum_{i=1}^{n} \xi_i(t)\boldsymbol{X}_i(t)n_i/\hat{\Lambda}_0(C_i)}{\sum_{i=1}^{n} \xi_i(t)n_i/\hat{\Lambda}_0(C_i)}$, 
$\overline{\hat{\boldsymbol{B}}}(t) = \frac{\sum_{i=1}^{n} \xi_i(t)\hat{\boldsymbol{B}}_i(t)n_i/\hat{\Lambda}_0(C_i)}{\sum_{i=1}^{n} \xi_i(t)n_i/\hat{\Lambda}_0(C_i)}$ are weighted averages of $\boldsymbol{X}_i(t)$ and $\hat{\boldsymbol{B}}_i(t)$, respectively, with weights corresponding to the relative number of visits. Here, $\hat{\Lambda}_0(t)$ is the Aalen-Breslow-type estimator of $\Lambda_0(t)$ \citep{liang2009joint}. Derivation details are provided in Supplementary Section S2. Note that time should be modeled in $\beta_0(t)$ rather than included in $\boldsymbol{X}_i(t)$ to avoid identification issues in the estimating equations for $\boldsymbol{\beta}$, as shown in Supplementary Section S3.

While these methods effectively address IP, they assume the biomarker is observed at every visit (i.e., $R_i(t)=1$ when $dN_i(t)=1$) or restrict the analysis to visits with recorded biomarker values. As showed later in the simulation studies, such exclusions can lead to biased estimates of the exposure effect.

\subsection{Our proposal}
\label{section_methods_EHRJoint}

\subsubsection{Model specification and assumptions}

Since JVML-Liang is the most flexible method under IP for capturing the dependencies between the visiting and longitudinal processes, we extend JMVL-Liang to account for both IP and IO, and introduce a tripartite joint model (EHRJoint) that simultaneously accounts for the visiting, observation, and longitudinal processes:
\begin{equation}\label{EHRJoint_VP}
    \mathbb{E}[dN_i(t)|\boldsymbol{W}_i,\eta_i] = \xi_i(t)\eta_i\exp(\boldsymbol{\gamma}^\top \boldsymbol{W}_i)d\Lambda_0(t);
\end{equation}
\begin{equation*}\label{EHRJoint_OP}
    logit\{\mathbb{P}(R_i(t)=1|dN_i(t)=1,\boldsymbol{V}_i)\} = \boldsymbol{\alpha}^\top \boldsymbol{V}_i;
\end{equation*}
\begin{equation}\label{EHRJoint_LP} (Y_i(t)|\boldsymbol{X}_i(t),\boldsymbol{b}_i) = \beta_0(t)+\boldsymbol{\beta}^\top \boldsymbol{X}_i(t) + \boldsymbol{b}_i^\top \boldsymbol{Z}_i(t)+\epsilon_i(t),
\end{equation}
where $\epsilon_i(t)$ is a zero mean measurement error process. The sub-model specification for the visiting process and the longitudinal process in \eqref{EHRJoint_VP} and \eqref{EHRJoint_LP} is the same as JMVL-Liang, and thus, the two processes are connected through both measured variables $\boldsymbol{W}_i$ and $\boldsymbol{X}_i(t)$, as well as the two random effects $\boldsymbol{b}_i$ and $\eta_i$. To account for IO, i.e., missing biomarker measurements, a logistic regression model is assumed for the observation process. Additionally, the following assumptions are required for coefficient estimation:

\begin{assumption} \label{EHRJoint_A1}
    Censoring time $C_i$ is noninformative in the sense that given covariates $(\boldsymbol{X}_i(t), \boldsymbol{W}_i,\boldsymbol{V}_i)$, $C_i$ is independent of the visiting process $N_i(t)$, the observation process $R_i(t)$, and the longitudinal outcomes $Y_i(t)$. 
\end{assumption}
\begin{assumption} \label{EHRJoint_A2}
    Given $\eta_i$ and covariates $\boldsymbol{W}_i, \boldsymbol{V}_i$, the visiting process and the observation process $(N_i(t),R_i(t))$ are independent of the longitudinal outcome $Y_i(t)$.
\end{assumption}
\begin{assumption} \label{EHRJoint_A3}
    The frailty parameter $\eta_i$ is assumed to follow a gamma distribution with mean $1$ and variance $\sigma_\eta^2$.
\end{assumption}
\begin{assumption} \label{EHRJoint_A4}
    The two random effects $\boldsymbol{b}_i$ and $\eta_i$ are assumed to satisfy the following linear relationship $ E(\boldsymbol{b}_i \mid \eta_i) = \boldsymbol{\theta} (\eta_i - 1)$, where $\boldsymbol{\theta}$ is a $q$-dimensional vector of parameters.
\end{assumption}

The conditional non-informative censoring assumption in Assumption \ref{EHRJoint_A1} is commonly made in longitudinal analysis. It implies that, given the observed covariates, the censoring time provides no additional information about the three processes $Y_i(t)$, $N_i(t)$, and $R_i(t)$. 
For example, in our data analysis presented in Section \ref{section_dataExample}, the assumption $Y_i(t) \perp\!\!\!\perp C_i \mid \boldsymbol{X}_i(t), \boldsymbol{W}_i, \boldsymbol{V}_i$ is satisfied because the study period is fixed at 5 years. This fixed duration represents administrative censoring, which is unrelated to the glucose level (our outcome). Therefore, the censoring time in this case can be considered non-informative.
The conditional independence assumption in Assumption \ref{EHRJoint_A2} enables us to decouple the modeling of the visiting and observation processes from the longitudinal process, so that parameters in the visiting and observation models can be estimated independently. In practice, the covariates $\boldsymbol{W}_i$ and $\boldsymbol{V}_i$ should include factors that both influence the biomarker values and drive patients to visit clinics and have their biomarkers measured, such as demographic and socioeconomic factors (e.g., age, gender), comorbidity conditions, and medication use. 
In Assumptions \ref{EHRJoint_A3} and \ref{EHRJoint_A4}, the parametric assumptions regarding the distribution of $\eta_i$ and the linear relationship between $\boldsymbol{b}_i$ and $\eta_i$ are made primarily for computational convenience. These assumptions are consistent with those in \cite{liang2009joint}. In addition, our simulation studies evaluate the robustness of our method when these assumptions are violated.

\subsubsection{Parameter estimation and inference}

We outline the key steps with details provided in Supplementary Section S4. Denote 
\newpage 
\noindent 
$\omega_i(t) = \mathbb{P}(R_i(t)=1|dN_i(t)=1,\boldsymbol{V}_i)$, $\mathcal{A}(t) = \int_0^t \beta_0(s) d\Lambda_0(s)$, and 
$\boldsymbol{B}_i(t) = \boldsymbol{Z}_i(t) \mathbb{E}[\boldsymbol{b}_i|n_i, C_i]= \boldsymbol{Z}_i(t) \mathbb{E}\big\{ (\eta_i - 1) \mid n_i, C_i \big\}$. 
All expectations here are implicitly conditional on $\boldsymbol{X}_i(t)$, $\boldsymbol{W}_i$, and $\boldsymbol{V}_i$. For simplicity and to avoid redundancy, we will omit $\boldsymbol{X}_i(t)$, $\boldsymbol{W}_i$, and $\boldsymbol{V}_i$ in the conditional expectations throughout the article. 
We first assume that $\omega_i(t), \Lambda_0(t), \boldsymbol{B}_i(t)$ are known or have been consistently estimated with $\widehat{\omega}_i(t), \widehat{\Lambda}_0(t), \widehat{\boldsymbol{B}}_i(t)$, when constructing the estimating equations for $\boldsymbol{\beta}$. We will discuss how to estimate these quantities later. 

With the primary goal of deriving the estimating equations for $\boldsymbol{\beta}$, we begin by taking the difference between the observed process of the outcome and its expected process:
\begin{equation*}
M_i(t, \boldsymbol{\beta}, \boldsymbol{\theta}) = \int_0^t [Y_i(s) - \boldsymbol{\beta}^\top \boldsymbol{X}_i(s) - \boldsymbol{\theta}^\top \widehat{\boldsymbol{B}}_i(s)] R_i(s)dN_i(s) 
- \int_0^t \xi_i(s) \widehat{\omega}_i(s) \frac{n_i}{\widehat{\Lambda}_0(C_i)} d\mathcal{A}(s).
\end{equation*}
In the observed process, the conditional expectation of the outcome involves unobserved random effects, which are marginalized out to account for their impact. The expected process has two components: one reflecting the visiting process of the patient and another corresponding to the observation process of the biomarker. Using $\mathbb{E}[M_i(t, \boldsymbol{\beta}, \boldsymbol{\theta})|n_i,C_i]=0$ and 
$\mathbb{E}[dM_i(t, \boldsymbol{\beta}, \boldsymbol{\theta})|n_i,C_i]=0$, we derive the following estimating equations for $\boldsymbol{\beta},\boldsymbol{\theta}$:
\begin{equation}\label{Estimating_eq_betatheta}
U_1(\boldsymbol{\beta}, \boldsymbol{\theta}) \equiv \sum_{i=1}^n \int_0^\tau 
\begin{pmatrix}
    \boldsymbol{X}_i(t) - \overline{\boldsymbol{X}}(t) \\
    \widehat{\boldsymbol{B}}_i(t) - \overline{\widehat{\boldsymbol{B}}}(t) 
\end{pmatrix}
\big\{ Y_i(t) - \boldsymbol{\beta}^\top \boldsymbol{X}_i(t) - \boldsymbol{\theta}^\top \widehat{\boldsymbol{B}}_i(t) \big\} R_i(t) dN_i(t) = 0, 
\end{equation}
where $\overline{\boldsymbol{X}}(t) = \frac{\sum_{j=1}^n \xi_j(t) \boldsymbol{X}_j(t) \widehat{\omega}_j(t) n_j / \widehat{\Lambda}_0(C_j)}{\sum_{j=1}^n \xi_j(t) \widehat{\omega}_j(t)  n_j / \widehat{\Lambda}_0(C_j)},\ 
\overline{\widehat{\boldsymbol{B}}}(t) = \frac{\sum_{j=1}^n \xi_j(t) \widehat{\boldsymbol{B}}_j(t) \widehat{\omega}_j(t) n_j / \widehat{\Lambda}_0(C_j)}{\sum_{j=1}^n \xi_j(t) \widehat{\omega}_j(t) n_j / \widehat{\Lambda}_0(C_j)}$.

Here $\overline{\boldsymbol{X}}(t)$ and $\overline{\widehat{\boldsymbol{B}}}(t)$ are weighted averages of $\boldsymbol{X}_i(t)$ and $\boldsymbol{B}_i(t)$, with weights involving the number of clinical visits and the probability of recording the biomarker to account for the impact from IP and IO, respectively.

Next, we present how to estimate $\omega_i(t), \Lambda_0(t), \boldsymbol{B}_i(t)$, as well as the key associated parameters $\boldsymbol{\gamma}$ and $\boldsymbol{\alpha}$. Firstly, $\boldsymbol{\gamma}$ is estimated using equation \eqref{EE_gamma_w}, which is the same as that in JMVL-Liang. $\Lambda_0(t)$ is consistently estimated by the Aalen-Breslow-type estimator:
\begin{align}
    \widehat{\Lambda}_0(t) &= \sum_{i=1}^n \int_0^t \frac{dN_i(s)}{\sum_{j=1}^n \xi_j(s) \exp(\widehat{\boldsymbol{\gamma}}^\top \boldsymbol{W}_j)}  \label{Estimating_Lambda}
\end{align}
Under the gamma distribution assumption for $\eta_i$ in Assumption \ref{EHRJoint_A3}, the expectation $\mathbb{E}(\eta_i \mid n_i, C_i)$ has a closed form \citep{liang2009joint}, and $\sigma_{\eta}^2$ and $\boldsymbol{B}_i(t)$ are estimated by
\begin{align}
    \widehat{\sigma}_{\eta}^2 &= min \bigg( \frac{\sum_{i=1}^n \big\{ n_i^2 - n_i - \exp(2 \widehat{\boldsymbol{\gamma}}^\top \boldsymbol{W}_i) \widehat{\Lambda}_0^2(C_i) \big\}}{\sum_{i=1}^n \exp(2 \widehat{\boldsymbol{\gamma}}^\top \boldsymbol{W}_i) \widehat{\Lambda}_0^2(C_i)}, \quad 0 \bigg),  \label{Estimating_sigma} \\
    \widehat{\boldsymbol{B}}_i(t) &= \frac{\big\{ n_i - \exp(\boldsymbol{\widehat{\gamma}}^\top \boldsymbol{W}_i) \widehat{\Lambda}_0(C_i) \big\} \widehat{\sigma}_{\eta}^2}{1 + \exp(\boldsymbol{\widehat{\gamma}}^\top \boldsymbol{W}_i) \widehat{\Lambda}_0(C_i) \widehat{\sigma}_{\eta}^2} \boldsymbol{Z}_i(t). \label{Estimating_Bhat}
\end{align}

Under a logistic regression model, $\boldsymbol{\alpha}$ is estimated using the following estimating equation:
\begin{equation}\label{Estimating_eq_alpha}
    U_2(\boldsymbol{\alpha}) = \sum_{i=1}^n \int_0^\tau  \boldsymbol{V}_i\big\{ R_i(t)- \frac{\exp(\boldsymbol{\alpha}^\top \boldsymbol{V}_i)}{1+\exp(\boldsymbol{\alpha}^\top \boldsymbol{V}_i)} \big\} dN_i(t)=0.
\end{equation}

Hence, with the resulting estimator $\boldsymbol{\widehat{\alpha}}$, we have $\widehat{\omega}_i(t)=\{1+\exp(\boldsymbol{\widehat{\alpha}}^\top \boldsymbol{V}_i)\}^{-1}\{\exp(\boldsymbol{\widehat{\alpha}}^\top \boldsymbol{V}_i)\}$.

We summarize the estimation procedure as follows:
\begin{algorithm}[H]
\caption{Estimation Procedure}
\begin{algorithmic}[1]
    \STATE Estimate $\boldsymbol{\gamma}$ using the estimating equation \eqref{EE_gamma_w}.
    \STATE Obtain $\widehat{\Lambda}_0(t)$ from \eqref{Estimating_Lambda}, $\widehat{\sigma}_\eta^2$ from \eqref{Estimating_sigma}, and $\widehat{\boldsymbol{B}}_i(t)$ from \eqref{Estimating_Bhat}.
    \STATE Estimate $\boldsymbol{\alpha}$ using the estimating equation \eqref{Estimating_eq_alpha}, and obtain $\widehat{\omega}_i(t)$.
    \STATE Estimate $\boldsymbol{\beta}$ and $\boldsymbol{\theta}$ using the estimating equation \eqref{Estimating_eq_betatheta}.
\end{algorithmic}
\end{algorithm}


Following \cite{liang2009joint}, we provide the theoretical properties of $(\widehat{\boldsymbol{\beta}}^\top,\widehat{\boldsymbol{\theta}}^\top)^\top$ estimated from the above estimating procedure. 

\begin{theorem}\label{theorem_1}
Suppose $\sigma_{\eta}^2 > 0$. Then we have
\begin{equation*}
    \sqrt{n}
    \begin{pmatrix}
        \widehat{\boldsymbol{\beta}} - \boldsymbol{\beta}\\
        \widehat{\boldsymbol{\theta}} -\boldsymbol{\theta}
    \end{pmatrix}
    \xrightarrow[]{d} \mathcal{N}(\boldsymbol{0}, \mathbf{H}^{-1} \mathbf{M} (\mathbf{H}^{-1})^\top)
\end{equation*}
as the sample size $n$ goes to infinity, where $\mathbf{H}$ 
and $\mathbf{M}$ are specified in the Supplementary Section S5.
\end{theorem}

The proof of Theorem \ref{theorem_1} is provided in Section S5 of the supplementary materials. The asymptotic variance–covariance matrix can be consistently estimated using the plug-in method. However, the analytic expression of $\mathbf{M}$ is complex, as it relies on the Taylor expansion of complex functions around the infinite-dimensional parameter $\Lambda_0(t)$. Consequently, direct estimation via the plug-in method may be unstable. Following the approach recommended by \cite{liang2009joint}, for practical applications we instead compute the variance using the bootstrap method.

\section{Simulation}
\subsection{Data generation}

\noindent \underline{\textbf{Setting A: Assessing impact of IP}}

Failing to account for IP can lead to biased parameter estimates. To assess the impact of IP, we conduct simulation studies using the methods reviewed in Section \ref{section_existing_methods} for estimating the exposure effect on the biomarker in the longitudinal model. 

We set the total number of patients to be $n$. For the $i$-th patient, we first generate two time-invariant variables, a binary exposure $A_i \sim Bernoulli(0.5)$ and a continuous variable $Z_i \sim \mathcal{N}(0,1)$. 
The clinical visit time $T_{ij}$ lies within the study period $(t_0, C_i]$, which is determined by the inter-visit time $S_{ij}$. The study period $(t_{0},C_i]$ is fixed, representing an administrative censoring time that is non-informative. 

We simulate $S_{ij}$ from six models, covering a spectrum from the most idealized to the most realistic EHR scenarios (Cases 1-1 to 1-6; see \textbf{Table \ref{table:simulation_cases}}):
Case 1-1 mimics classical longitudinal studies with constant inter-visit times. In Case 1-2, $S_{ij}$ is determined by measured covariates. Case 1-3 introduces additional complexity, with $S_{ij}$ depending on a latent variable that correlates with the random effects in the longitudinal model. Cases 1-4 and 1-5 further incorporate dependencies on the biomarker value at the previous visit (Case 1-4) or the current biomarker value (Case 1-5). Finally, Case 1-6 simulates a mixture of patterns from Cases 1-1 to 1-5. Given visit time $T_{ij}$, we generate the biomarker value $Y_{ij}$ from a linear mixed-effects model. The data-generating model is:
\begin{align*}
    (S_{ij}=T_{ij}-T_{i(j-1)}) &\sim Exponential(\lambda_{ij}), \lambda_{ij}=\eta_{i}\exp(\gamma_0+\gamma_a A_i + \gamma_z Z_i + \gamma_u U_{ij}), \\
    (Y_{ij}|A_i,Z_i,b_{0i},b_{1i},T_{ij}) &= (\beta_0 + b_{0i})  + (\beta_a+b_{1i}) A_i + \beta_z Z_i + \beta_t T_{ij} + \epsilon_{ij},
\end{align*}
and $\epsilon_{ij} \sim \mathcal{N}(0,\sigma_{\epsilon}^2)$, and random effects $\boldsymbol{b}_i = (b_{0i},b_{1i})^\top \sim MVN(\boldsymbol{0}, \Sigma_b)$. Variable $U_{ij}$ is a generic placeholder that varies across simulation cases. For example, $U_{ij}$ is NULL in Cases 1-2 and 1-3, the previous biomarker value in Case 1-4, and the current value in Case 1-5.
Further parameter details are provided in Supplementary Section S6.
Our primary focus is coefficient estimation in the longitudinal model, particularly the binary exposure effect $\beta_a$.

\noindent \underline{\textbf{Setting B: Assessing joint impact of IP and IO}}

In situations where the biomarker is not observed at each visit, we simulate data reflecting both IP and IO. We construct three simulation cases summarized in \textbf{Table \ref{table:simulation_cases}} (Case 2-1 to Case 2-3). Case 2-1 represents a scenario without IP or IO. Case 2-2 reflects a setting with IP but without IO. Case 2-3 is under both IP and IO. We simulate data using the following joint model, where a key difference of the data generation models between Setting B and Setting A lies in the observation process model for the biomarker:
\begin{align}
    \mathbb{E}[dN_i(t)|A_i, Z_i,\eta_i] &= \xi_i(t)\eta_i\exp(\gamma_a A_i+\gamma_z Z_i)d\Lambda_0(t), \nonumber\\
    logit\{\mathbb{P}(R_i(t)=1 &| dN_i(t)=1,A_i, Z_i)\} = \alpha_0+ \alpha_a A_i + \alpha_z Z_i, \nonumber \\
    (Y_i(t)|A_i, Z_i,b_{0i},b_{1i}) &= (\beta_0+b_{0i}) + (\beta_a+b_{1i})A_i + \beta_z Z_i + \beta_t t +\epsilon_i(t), \nonumber
\end{align}
\noindent where $d\Lambda_0(t)=1$, and $\eta_i$ follows a gamma distribution with mean 1 and variance $\sigma_\eta^2$. Random effects $(b_{0i}, b_{1i})^\top \sim \mathcal{N}(\boldsymbol{\theta}(\eta_i - 1), \Sigma_b)$, and $\epsilon_i(t) \sim \mathcal{N}(0, \sigma_\epsilon^2)$.

\noindent \underline{\textbf{Setting C: Evaluation under model misspecification}}

In contrast to Setting B, where our proposed method, EHRJoint, is evaluated under correct model specification, this section evaluates our method and competing methods under conditions of model misspecification, with a focus on misspecification of the visiting process model. Specifically, we generate visit times using the same six visiting process models as in Setting A. However, the visiting process model is fitted only according to the specification in Setting B and thus is misspecified. The observation indicator $R_i(t)$ and the longitudinal biomarker $Y_i(t)$ is generated using the same models as in Setting B, so these components remain correctly specified. Although the observation process model could also be misspecified, our focus here is on the visiting process model, as it often poses greater challenges for correct model specification due to the difficulty of collecting all factors that impact clinical visits.

\begin{table}[ht]
\caption{Description of simulation cases.}
\label{table:simulation_cases}
\resizebox{1\textwidth}{!}{
\begin{tabular}{c>{\raggedright}p{4cm}>{\raggedright}p{13cm}}
\hline 
\textbf{Case} & \textbf{Short Description} & \textbf{Details}\tabularnewline
\hline 
\multicolumn{3}{c}{\textbf{Assessing impact of IP}}\tabularnewline
\hline 
1-1 & Regular visits (non-IP) & The inter-visit time is constant, i.e., $S_{ij}=c$, so the visiting
process is uninformative.\tabularnewline
\hline 
1-2 & Shared measured variables & The visiting process model depends on measured variables, which are
shared with the longitudinal model. We generate $S_{ij}$ from an
exponential distribution with intensity $\lambda_{i}=\exp\big(\gamma_{0}+\gamma_{a}A_{i}+\gamma_{z}Z_{i}\big)$. \tabularnewline
\hline 
1-3 & Dependent latent variables & The visiting process model depends on latent variables, which are
associated with random effects in the longitudinal model. We generate
$S_{ij}$ from an exponential distribution with intensity $\lambda_{i}=\eta_{i}\exp\big(\gamma_{0}+\gamma_{a}A_{i}+\gamma_{z}Z_{i}\big)$,  $\eta_{i}|\boldsymbol{b}_{i}\sim Gamma(\exp(\gamma_{b}b_{1i})^{-1},\sigma_{\eta}^{2})$.\tabularnewline
\hline 
1-4 & Previous biomarker driven & The visiting process depends on the previous biomarker value. We generate
$S_{ij}$ from an exponential distribution with intensity $\lambda_{ij}=\exp\big(\gamma_{0}+\gamma_{a}A_{i}+\gamma_{z}Z_{i}+\gamma_{y}y_{i(j-1)}\big)$.\tabularnewline
\hline 
1-5 & Current biomarker threshold & The visiting process model depends on the current biomarker value.
Visits occur when biomarker values exceed a pre-specified threshold
(top 20\%). This simulates patients visiting when biomarker levels
reach \textquotedbl abnormal\textquotedbl{} levels.\tabularnewline
\hline 
1-6 & Mixed patterns & A mixture of the above five processes with equal probability.\tabularnewline
\hline 
\multicolumn{3}{c}{\textbf{Assessing joint impact of IP and IO}}\tabularnewline
\hline 
2-1 & Regular visits (non-IP) and non-IO & The inter-visit time is constant, i.e., $S_{ij}=c$. The observation
process is non-informative with $\boldsymbol{\alpha}=\boldsymbol{0}$.\tabularnewline
\hline 
2-2 & IP and non-IO & Simulate $S_{ij}$ corresponding to the model $\mathbb{E}[dN_{i}(t)|A_{i},Z_{i},\eta_{i}]=\xi_{i}(t)\eta_{i}\exp(\gamma_{a}A_{i}+\gamma_{z}Z_{i})d\Lambda_{0}(t)$.
The observation process is non-informative with $\boldsymbol{\alpha}=\boldsymbol{0}$.\tabularnewline
\hline 
2-3 & IP and IO & Simulate $S_{ij}$ corresponding to the model $\mathbb{E}[dN_{i}(t)|A_{i},Z_{i},\eta_{i}]=\xi_{i}(t)\eta_{i}\exp(\gamma_{a}A_{i}+\gamma_{z}Z_{i})d\Lambda_{0}(t)$.
Simulate IO with $\boldsymbol{\alpha}=(-2,2,1)$.\tabularnewline
\hline 
\end{tabular}
}
\end{table}


\subsection{Fitted approaches}\label{simulation_fittedMethods}

We include the following methods for comparison:
(1) The summary statistics approach used in LabWAS based on the min, mean, median, max of the longitudinal measurements. 
(2) Mixed-effects models, including a standard linear mixed-effects model (Standard LME), LME conditioning on the historical number of biomarker observations (OA-LME), and LME conditioning on the historical number of visits (VA-LME).
(3) Joint models of visiting and longitudinal processes, including JMVL-LY, IIRR-weighting, JMVL-G, and JMVL-Liang (see Section \ref{section_existing_methods}).
(4) Joint models of visiting, observation, and longitudinal processes, including our proposed method EHRJoint, and a modified JMVL-Liang with a logistic model for the observation process and coefficients fixed at $\boldsymbol{0}$ (Adapted-Liang).

In Setting A, where we evaluate only the impact of IP, and IO is not involved, OA-LME and VA-LME are identical because the number of visits equals that of biomarker measurements. Additionally, EHRJoint and Adapted-Liang reduce to JMVL-Liang. As a result, these methods are excluded from the comparison in Setting A.

\subsection{Results}
We present results of the exposure effect $\beta_a$ estimates in the longitudinal model based on 500 replications. In IP-only settings, JMVL-Liang yields the least bias among existing methods. Under both IP and IO, EHRJoint is the only method that remains unbiased. EHRJoint also performs the best with the smallest bias even under model misspecification across all scenarios with various models for IP and IO.

\noindent \underline{\textbf{Results of Setting A: assessing the impact of IP}}

\textbf{Bias.} \textbf{Table \ref{sim_results_settingA}} presents results of $\beta_a$, the effect of exposure $A$ in the longitudinal model. In the absence of IP (Case 1-1), all methods are nearly unbiased. 
In Case 1-2, where the visiting and longitudinal processes depend only on measured variables, most methods remain unbiased, except those using minimum and maximum summary statistics. Under other IP scenarios, the summary-statistic approach produces biased estimates, with direction and magnitude varying by the chosen statistic.
In Case 1-3, where unmeasured random effects in the visiting and longitudinal models are dependent, noticeable bias arises across all methods. Among methods using repeated measurements, JMVL-LY has the largest bias (-0.81, SD: 0.29), followed by IIRR-weighting (-0.79, SD: 0.18) and JMVL-G (-0.61, SD: 0.27). In contrast, JMVL-Liang, which incorporates dependent latent variables, shows the smallest bias (-0.19, SD: 0.25). Nonetheless, JMVL-Liang does not completely remove bias in $\hat{\beta}_a$, likely due to misspecification of the frailty parameter distribution.
In more complex IP scenarios, such as when the visiting process depends on the previous biomarker (Case 1-4), the current biomarker (Case 1-5), or mixed patterns (Case 1-6), all methods exhibit bias in estimating $\beta_a$, and JMVL-Liang produces the smallest bias among all.

\textbf{Variance.} The Standard LME and OA-LME models, which use repeated measurements, have lower variance in $\hat{\beta}_a$ than summary-statistic methods, likely due to their larger sample size. For example in Case 1-3, the standard deviation of $\hat{\beta}_a$ is 0.13 from the Standard LME, compared to 0.16 from the mean summary statistic approach. On the other hand, JMVL-LY, 
\newpage 
\noindent 
JMVL-G, JMVL-Liang, and the IIRR-weighting method exhibit higher variance due to additional variability introduced by the visiting process (JMVL-LY: 0.29, JMVL-G: 0.27, JMVL-Liang: 0.25, IIRR-weighting: 0.18). 

\begin{table}[H]
\centering
\scriptsize
\caption{Results of bias , standard deviation (SD), and root mean squared error (RMSE) under simulation Setting A to evaluate the performance of existing methods only under IP. Values are multiplied by 100. Values with the smallest absolute bias, SD, and RMSE are bolded. $^\ast$: When only IP is involved, VA-LME and OA-LME are identical, and both EHRJoint and Adapted-Liang reduce to JMVL-Liang. OA-LME fails to converge in Case 1-1 due to perfect correlation (r = 1) between time and the number of visits.}\label{sim_results_settingA}
\renewcommand{\arraystretch}{1.2}
\resizebox{\textwidth}{!}{
\begin{tabularx}
{\columnwidth}{msssssss}
\hline
Method & Metric & Case 1-1 & Case 1-2 & Case 1-3 & Case 1-4 & Case 1-5 & Case 1-6 \\
 & & Non-IP & Share measured variables & Dependent latent variables & Previous biomarker driven & Current biomarker threshold & Mixed patterns \\
\midrule
\multirow{3}{*}{Min} 
& Bias   & \textbf{0.5}      & -49.5    & -98.3    & 71.4     & 51.5     & -28.5    \\
& SD     & 11.9     & 13.5     & 19.6     & 13.0       & \textbf{1.4}      & 19.1     \\
& RMSE   & 11.9     & 51.3     & 100.2    & \textbf{72.6}     & 51.5     & 34.3     \\
\hline
\multirow{3}{*}{Mean} 
& Bias   & 0.8      & \textbf{0.7}      & -26.5    & 100.1    & 67.0      & 6.1      \\
& SD     & \textbf{11.2}     & 12.3     & 16.1     & 18.7     & 4.0        & 17.9     \\
& RMSE   & \textbf{11.2}     & 12.3     & \textbf{31.0}       & 101.9    & 67.1     & \textbf{18.9}     \\
\hline
\multirow{3}{*}{Median} 
& Bias   & 0.9    & 0.8    & -26.5  & 99.5   & 68.0   & 6.6 \\
& SD     & 11.5   & 13.0   & 16.7   & 19.3   & 4.1    & 18.1 \\
& RMSE   & 11.5   & 13.0   & 31.3   & 101.4  & 68.1   & 19.3 \\
\hline
\multirow{3}{*}{Max} 
& Bias   & 0.6      & 50.7     & 45.7     & 129.8    & 85.8     & 39.5     \\
& SD     & 12.2     & 13.2     & 17.3     & 27.6     & 10.3     & 19.9     \\
& RMSE   & 12.2     & 52.4     & 48.8     & 132.7    & 86.4     & 44.2     \\
\hline
\multirow{3}{*}{Standard LME} 
& Bias   & 0.8      & \textbf{0.7}      & -28.2    & 107.7    & 76.2     & 18.9     \\
& SD     & \textbf{11.2}     & \textbf{11.3}     & \textbf{13.4}     & 12.2     & 5.6      & \textbf{16.0}       \\
& RMSE   & \textbf{11.2}     & \textbf{11.3}     & 31.2     & 108.4    & 76.4     & 24.7     \\
\hline
\multirow{3}{*}{OA-LME$^\ast$} 
& Bias   & -      & \textbf{0.7}      & -28.1    & 104.8    & 53.7     & 20.5     \\
& SD     & -     & \textbf{11.3}     & \textbf{13.4}     & \textbf{12.0}       & 2.0        & \textbf{16.0}       \\
& RMSE   & -     & \textbf{11.3}     & 31.2     & 105.4    & 53.7     & 26.0       \\
\hline
\multirow{3}{*}{JMVL-LY} 
& Bias   & 0.8      & 1.0        & -80.5    & 160.6    & 109.3    & 51.1     \\
& SD     & \textbf{11.2}     & 13.4     & 29.4     & 12.6     & 7.6      & 21.5     \\
& RMSE   & \textbf{11.2}     & 13.5     & 85.7     & 161.0      & 109.6    & 55.4     \\
\hline
\multirow{3}{*}{IIRR-weighting} 
& Bias   & 0.8      & \textbf{0.7}      & -78.9    & 168.2    & 114.0      & 70.0       \\
& SD     & \textbf{11.2}     & 11.9     & 18.3     & 12.8     & 7.6      & 20.3     \\
& RMSE   & \textbf{11.2}     & 11.9     & 81.0       & 168.7    & 114.2    & 72.8     \\
\hline
\multirow{3}{*}{JMVL-G} 
& Bias   & 1.7      & 0.8      & -60.7    & 214.5    & 141.2    & 70.9     \\
& SD     & 15.1     & 14.2     & 26.6     & 32.8     & 19.7     & 36.6     \\
& RMSE   & 15.2     & 14.2     & 66.3     & 217.0      & 142.5    & 79.8     \\
\hline
\multirow{3}{*}{JMVL-Liang$^\ast$} 
& Bias   & 0.8      & 0.9      & \textbf{-18.8}    & \textbf{96.5}     & \textbf{33.2}     & \textbf{5.4}      \\
& SD     & \textbf{11.2}     & 13.9     & 25.0       & \textbf{12.0}       & 6.4      & 26.3     \\
& RMSE   & \textbf{11.2}     & 13.9     & 31.3     & 97.2     & \textbf{33.8}     & 26.9   \\ 
\hline
\end{tabularx}}
\end{table}

Bias, variance, and MSE for other coefficient estimates, $\hat{\beta}_0$, $\hat{\beta}_z$, and $\hat{\beta}_t$, in the longitudinal model are presented in \textbf{Tables S1-S3}. JMVL-Liang and JMVL-LY do not estimate $\beta_0$ and $\beta_t$ due to identifiability issues. In Case 1-3, the Standard LME produces nearly unbiased estimates for $\beta_0$, $\beta_z$, and $\beta_t$, since only variable $A$ has random effects shared with the visiting process. Consequently, the visiting process has minimal impact on coefficient estimates other than $\hat{\beta}_a$ \citep{mcculloch2016biased}. Additionally, in this case, all methods show small bias and MSE for $\hat{\beta}_z$, except the minimum and maximum summary statistics approach. 

\noindent \underline{\textbf{Results of Setting B: assessing the joint impact of IP and IO}}

\begin{figure}[H]
    \centering
    \includegraphics[width=1\linewidth]{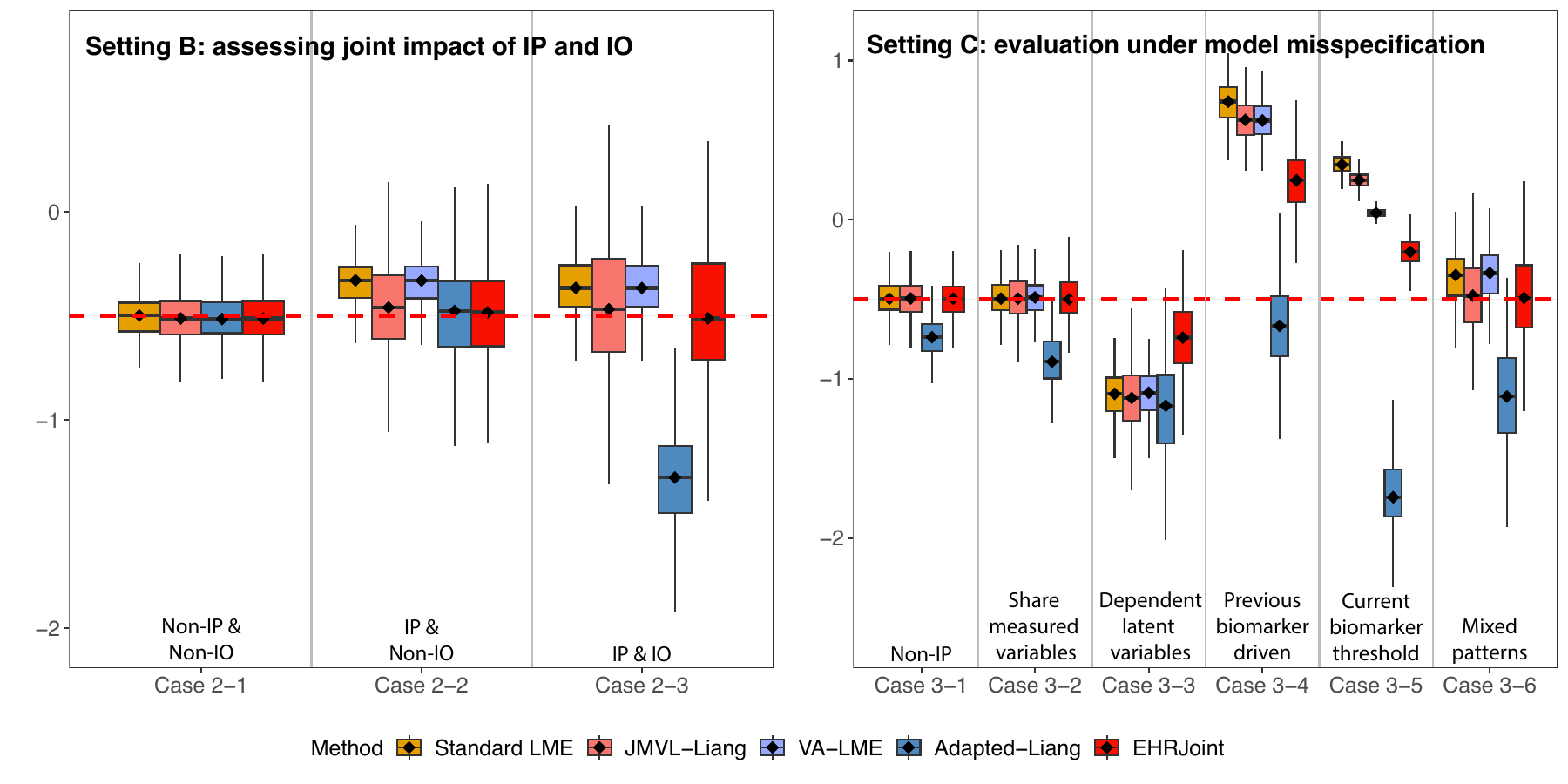}
    \caption{Box plot of the estimated exposure effect of $A$ on the longitudinal biomarker based on 500 simulation replications under simulation settings B and C. Both settings involve the presence of IP and IO. 
    Setting B evaluates the joint impact of IP and IO under three cases where sub-models in EHRJoint are correctly specified: Case 2-1 (non-IP and non-IO), Case 2-2 (IP but non-IO), and Case 2-3 (IP and IO). In contrast, Setting C (Cases 3-1 to 3-6) examines the performance under visiting process model misspecification. We simulate six scenarios representing a spectrum of EHR visiting patterns with increasing dependency complexity. The dotted horizontal line represents the true value of the coefficient.}
    \label{fig:simulation_settingBC_plot}
\end{figure}

\textbf{Figure \ref{fig:simulation_settingBC_plot}} and \textbf{Table \ref{sim_results_settingBC_table}} show results from a selected subset of methods, including Standard LME, VA-LME, JMVL-Liang, Adapted-Liang and EHRJoint. Full results are presented in \textbf{Figure S1(B)} and \textbf{Tables S4-S6}. In the absence of IP and IO (Case 2-1), all methods are nearly unbiased. However, in all other cases, EHRJoint is the only method that remains nearly unbiased, whereas Standard LME has the largest bias. 
Under IP and IO (Case 2-3), the bias for EHRJoint is $1.7\times10^{-2}$ (SD: 0.36), compared to biases that are 2.4, 8.3, and 46.9 times larger for JMVL-Liang, Standard LME, and Adapted-Liang, respectively.
While EHRJoint yields the smallest bias, it has the largest variance, due to additional variability introduced by modeling both the visiting and observation processes. In Case 2-3, the standard deviation is 0.36 for EHRJoint, 0.34 for JMVL-Liang, and 0.16 for Standard LME.
\vspace{-0.5 cm}

\noindent \underline{\textbf{Results of Setting C: evaluation under model misspecification}}

\textbf{Figure \ref{fig:simulation_settingBC_plot}} and \textbf{Table \ref{sim_results_settingBC_table}} present results under simulation Setting C involving both IP and IO, with potential misspecification of the IP model. Full results of all methods are available in \textbf{Figure S1 (C)} and \textbf{Tables S7-S9}. Similar to the IP-only scenarios, all methods are nearly unbiased in the absence of IP and IO (Case 3-1) or when the visiting and longitudinal processes only share measured variables (Case 3-2). In complex scenarios (Case 3-3 to Case 3-6) with both IP and IO, EHRJoint consistently exhibits the smallest bias among all methods, despite misspecifying the IP model across all approaches. In contrast, JMVL-Liang, which has the smallest bias in the IP-only scenarios, exhibits large bias due to ignoring IO. 
\vspace{-0.5cm}
\begin{table}[H]
\centering
\scriptsize
\caption{Results of bias, standard deviation (SD), and root mean squared error (RMSE) under simulation settings B and C across a subset of methods. Values are multiplied by 100. Both settings involve the presence of IP and IO. Setting B evaluates the joint impact of IP and IO, while Setting C examines performance under IP model misspecification. Values with the smallest absolute bias, SD, and RMSE are bolded. VA-LME fails to converge in Cases 2-1 and 3-1 due to perfect correlation (r = 1) between time and the number of visits. } \label{sim_results_settingBC_table}
\renewcommand{\arraystretch}{1.2}
\begin{tabularx}
{\columnwidth}{so|sss|ssssss}
\hline
& & \multicolumn{3}{c|}{\shortstack{Setting B: \\Assessing joint impact of IP and IO}} & \multicolumn{6}{c}{\shortstack{Setting C: \\ Evaluation under model misspecification}} \\ \hline
Method & Metric & Case 2-1 & Case 2-2 & Case 2-3 & Case 3-1 & Case 3-2 & Case 3-3 & Case 3-4 & Case 3-5 & Case 3-6 \\
& & \Shortunderstack{Non-IP\\ \& Non-IO} & \Shortunderstack{IP \& \\ Non-IO} & \Shortunderstack{IP\&IO} & \Shortunderstack{Non-IP} & \Shortunderstack{Share \\ measured \\ variables} & \addstackgap{\Shortunderstack{Dependent \\ latent \\ variables}} & \Shortunderstack{Previous \\ biomarker \\ driven} & \Shortunderstack{Current \\ biomarker \\ threshold} & \Shortunderstack{Mixed \\ patterns} \\
\hline
\multirow{3}{*}{\shortstack{Standard\\LME}} 
  & Bias & \textbf{-0.5} & 16.3 & 14.1 & 0.8 & \textbf{0.7} & -59.6 & 124.2 & 84.9 & 13.7 \\
  & SD   & \textbf{10.5} & \textbf{12.8} & 15.8 & \textbf{11.7} & \textbf{11.7} & 15.9 & 12.7 & 6.2 & \textbf{17.0} \\
  & RMSE & \textbf{10.5} & 20.7 & 21.2 & \textbf{11.7} & \textbf{11.8} & 61.6 & 124.8 & 85.2 & \textbf{21.8} \\
\hline
\multirow{3}{*}{\shortstack{JMVL-\\Liang}} 
  & Bias & -1.1 & 4.0 & 4.1 & 0.7 & \textbf{0.7} & -62.7 & 112.6 & 74.8 & 2.9 \\
  & SD   & 11.8 & 23.9 & 33.5 & 13.1 & 14.5 & 22.6 & 12.9 & 5.1 & 26.2 \\
  & RMSE & 11.8 & 24.2 & 33.7 & 13.1 & 14.5 & 66.6 & 113.4 & 75.0 & 26.4 \\
\hline
\multirow{3}{*}{VA-LME} 
  & Bias & - & 16.1 & 14.2 & - & 0.9 & -59.0 & 112.4 & 54.1 & 15.5 \\
  & SD   & - & 12.9 & \textbf{15.0} & - & 11.8 & \textbf{15.8} & \textbf{11.6} & \textbf{2.7} & 17.1 \\
  & RMSE & - & \textbf{20.6} & \textbf{20.7} & - & 11.9 & 61.0 & 113.0 & 54.2 & 23.0 \\
\hline
\multirow{3}{*}{\shortstack{Adapted-\\Liang}} 
  & Bias & -0.9 & \textbf{1.5} & -79.8 & -23.2 & -38.4 & -68.3 & \textbf{-17.1} & -122.8 & -61.6 \\
  & SD   & 12.3 & 24.6 & 23.9 & 14.1 & 16.4 & 30.9 & 28.3 & 21.7 & 33.6 \\
  & RMSE & 12.4 & 24.6 & 83.3 & 27.2 & 41.7 & 75.0 & \textbf{33.0} & 124.8 & 70.1 \\
\hline
\multirow{3}{*}{EHRJoint} 
  & Bias & -1.1 & \textbf{1.5} & \textbf{1.7} & \textbf{0.5} & \textbf{0.7} & \textbf{-25.1} & 74.5 & \textbf{30.0} & \textbf{1.9} \\
  & SD   & 11.8 & 24.3 & 35.6 & 13.1 & 14.2 & 23.9 & 20.0 & 10.4 & 30.8 \\
  & RMSE & 11.8 & 24.4 & 35.6 & 13.1 & 14.2 & \textbf{34.7} & 77.1 & \textbf{31.7} & 30.9 \\
\hline
\end{tabularx}
\end{table}

\vspace{-1cm}

\section{Analysis of the Michigan Genomics Initiative data} \label{section_dataExample}
\subsection{Data sources, preprocessing, and analytical sample}

We apply the methods described in Section \ref{simulation_fittedMethods} to analyze longitudinal glucose measurements using data from the MGI. We focus on two exposures: (1) genetic variants (SNPs) and (2) educational disadvantage. Due to significant differences in the lab measurement patterns between in-patient and out-patient visits, e.g., in-patient visits are high frequency and intrinsically different in nature, we restrict the analysis to out-patient visits only. We further select a cohort of participants aged $\geq18$ at enrollment who have at least five-year follow-up. The study period is defined as the first five years since enrollment. 

For the genetic analysis, we examine five SNPs previously identified as significant in LabWAS \citep{goldstein2020labwas} and reported in the GWAS catalog: rs7903146, rs10811661, rs3802177, rs7651090, and rs780094. To be consistent with LabWAS, we include all eligible patients regardless of disease or medication status. Furthermore, we restrict the analysis to individuals of European ancestry to minimize the impact of population structure.
To study the association with educational disadvantage, we use a neighborhood-level measure (proportion of adults with less than high school diploma) from the \href{https://nanda.isr.umich.edu/}{National Neighborhood Data Archive} and record it into quartiles to preserve privacy \citep{beesley2020emerging}.

We include the following covariates in the respective components of the joint model:
\begin{itemize}
\item \textbf{Visiting process}: Age at enrollment, sex, marital status, educational disadvantage, cancer status, chronic disease status, SNP (only in the SNP analysis), and race/ethnicity (only in the educational disadvantage analysis)
\item \textbf{Observation process}: covariates in the visiting process model, baseline BMI, and pancreatic disease status. 
\item \textbf{Longitudinal process}: Age at enrollment, sex, and the SNP of interest for SNP analysis; Age at enrollment, sex, race/ethnicity, marital status, educational disadvantage, cancer status, chronic disease status, baseline BMI, and pancreatic disease status for educational disadvantage analysis.
\end{itemize}

\begin{figure}
    \centering
    \includegraphics[width=1\linewidth]{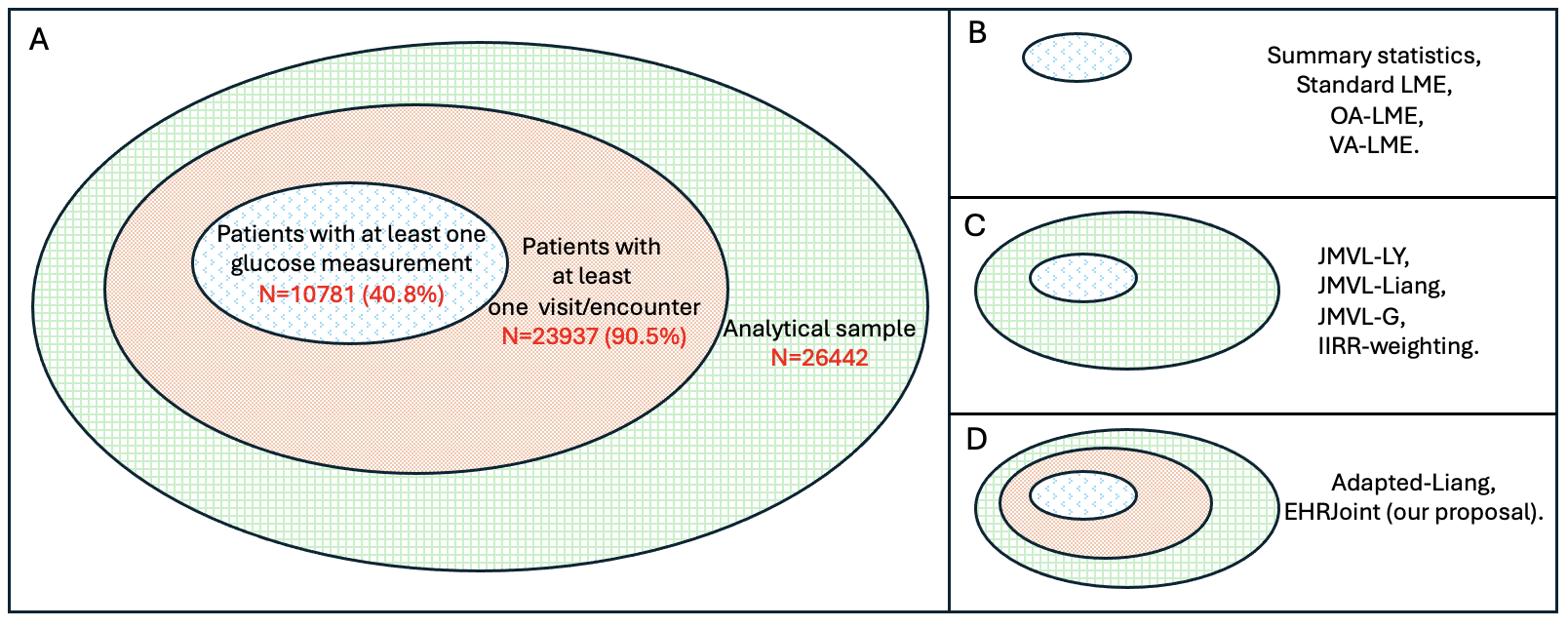}
    \caption{    
    Categorization of patients in the Michigan Genomics Initiative (A) and the subsets of data used by each method (B-D). The study period is defined as the first five years since patient enrollment. Around 9.5\% participants in the study have fully measured baseline covariates but lack outpatient visit records.}
    \label{fig:dataExample_vennDiag}
\end{figure}

\textbf{Figure \ref{fig:dataExample_vennDiag}} illustrates the construction of the analytical sample ($N=26,442$) from the MGI and the descriptive characteristics of these patients are shown in \textbf{Table \ref{tableone}}. 
Patients are categorized into three groups based on available data: those with no visit information or 
\newpage 
\noindent 
glucose measurements, those with at least one visit or encounter ($N=23,937$, 90.5\%), and those with at least one glucose measurement ($N=10,781$, 40.8\%). Different methods rely on different subsets of this data. Methods such as the summary statistics approach using minimum, mean, median or maximum of the longitudinal measurements, Standard LME, OA-LME, and VA-LME focus exclusively on patients with glucose measurements. Others, including JMVL-LY, JMVL-Liang, JMVL-G, and IIRR-weighting, utilize data from all patients but do not distinguish between patients with visits and those without. Adapted-Liang and EHRJoint incorporate all three groups of patients, fully leveraging information on both visits and glucose measurements.

\subsection{Results}

\noindent \underline{\textbf{Summary of the visiting process model}}

Researchers may also be interested in understanding which factors influence how frequently patients visit clinics, as well as the magnitude of these effects. These associations are captured by the estimated coefficients $\widehat{\boldsymbol{\gamma}}$ in the visiting process model (\textbf{Figures S2}). In the SNP analysis, EHRJoint identifies sex, marital status, and the presence of cancer or chronic disease as significant predictors of visit rates. As expected, disease indicators are associated with more frequent clinic visits: patients with cancer have a 2.11 (2.06, 2.15)-fold increase in visit rate compared to those without cancer, holding other factors constant; similarly, those with chronic diseases show a 3.32 (3.24, 3.40)-fold increase in visit rate. Interestingly, being married is associated with a 10.1\% (7.6\%-12.5\%)  decrease in the clinic visit rate compared to unmarried individuals, adjusting for other covariates. 
While this finding may seem counterintuitive, since marriage is often associated with greater social support and healthcare access, it is important to note that our ``unmarried'' reference group consists of not only single individuals but also those who are widowed or divorced. These subgroups may have increased visit rates due to greater health needs.

\vspace{-0.6cm}

\begin{table}[H]
\centering
\scriptsize
\caption{Descriptive characteristics of MGI patients included in the analysis. Continuous variables are summarized using means and standard deviations; categorical variables are presented as counts and percentages.}\label{tableone}
\begin{tabular}{lcccc} 
\hline                                                                                     & All            & \begin{tabular}[c]{@{}c@{}}Patients with \\ at least one visit\\ and one measurement\end{tabular} & \begin{tabular}[c]{@{}c@{}}Patients with \\ at least one visit\end{tabular} & \begin{tabular}[c]{@{}c@{}}Patients with \\ at least one visit\\ but no measurement\end{tabular} \\ \hline
Number of Patients                                                                           & 26442          & 10781                                                                                             & 23937                                                                       & 13156                                                                                            \\
Age (Mean (SD))                                                                              & 46.3 (13.9)    & 49.2 (14.7)                                                                                       & 46.5 (13.9)                                                                 & 44.2 (12.8)                                                                                      \\
Baseline BMI (Mean (SD))                                                                     & 30.1 (6.8)     & 30.4 (7.0)                                                                                          & 30.1 (6.8)                                                                  & 29.9 (6.7)                                                                                       \\
Educational disadvantage (Mean (SD))                                                         & 1.7 (0.8)      & 1.8 (0.8)                                                                                         & 1.7 (0.8)                                                                   & 1.6 (0.8)                                                                                        \\
Sex = Male (\%)                                                                              & 11876 (44.9)   & 5078 (47.1)                                                                                       & 10698 (44.7)                                                                & 5620 (42.7)                                                                                      \\
Marital status = 1 (\%)                                                                      & 18521 (70.0)     & 7393 (68.6)                                                                                       & 16751 (70.0)                                                                  & 9358 (71.1)                                                                                      \\
Cancer = 1 (\%)                                                                              & 11214 (42.4)   & 6795 (63.0)                                                                                         & 11214 (46.8)                                                                & 4419 (33.6)                                                                                      \\
Chronic disease =1 (\%)                                                                      & 17177 (65.0)     & 9453 (87.7)                                                                                       & 17177 (71.8)                                                                & 7724 (58.7)                                                                                      \\
Pancreatic disease =1 (\%)                                                                   & 3854 (14.6)    & 2809 (26.1)                                                                                       & 3854 (16.1)                                                                 & 1045 (7.9)                                                                                       \\
\begin{tabular}[c]{@{}l@{}}Number of Visits \\ (min, mean, median, max)\end{tabular}         & (0,11.5,8,190) & (1,19.3,15,190)                                                                                   & (1,12.7,9,190)                                                              & (1,7.3,5,90)                                                                                     \\
\begin{tabular}[c]{@{}l@{}}Number of Observations \\ (min, mean,   median, max)\end{tabular} & (0,2.5,0,162)  & (1,6.1,3,162)                                                                                     & (0,2.7,0,162)                                                               & (0,0,0,0)                                                                                        \\
Minor allele frequency                                                                       &                &                                                                                                   &                                                                             &                                                                                                  \\
\quad rs7903146                                                               & 0.289          & 0.286                                                                                             & 0.289                                                                       & 0.292                                                                                            \\
\quad rs10811661                                                              & 0.171          & 0.170                                                                                              & 0.171                                                                       & 0.171                                                                                            \\
\quad rs3802177                                                               & 0.302          & 0.301                                                                                             & 0.302                                                                       & 0.302                                                                                            \\
\quad rs7651090                                                               & 0.317          & 0.317                                                                                             & 0.317                                                                       & 0.318                                                                                            \\
\quad rs780094                                                                & 0.407          & 0.408     & 0.407                                                                       & 0.406   \\ \hline                                                                            
\end{tabular}
\end{table}

\noindent \underline{\textbf{Summary of the observation process model}} 

Estimated coefficients from the EHRJoint observation process model are presented in \textbf{Figure S3}. Significant predictors include age, sex, race/ethnicity, educational disadvantage, cancer, chronic disease, pancreatic disease, and baseline BMI. Notably, a one-quartile increase in educational disadvantage is associated with a 1.18 (1.14, 1.22)-fold increase in the odds of glucose being recorded, after adjusting for other covariates. 
One explanation is that educational disadvantage may correlate with other unmeasured patient characteristics, such as health literacy and medication adherence, which could influence provider decisions to monitor glucose levels more closely \citep{ilhan2021health}.
Results for the educational disadvantage analysis are similar to the SNP analysis and are therefore not repeated here (\textbf{Figures S4-S5}).

\noindent \underline{\textbf{Summary of the exposure-biomarker association}}

\textbf{(A) SNP-glucose association.} 

\begin{figure}
    \centering
    \includegraphics[width=1\linewidth]{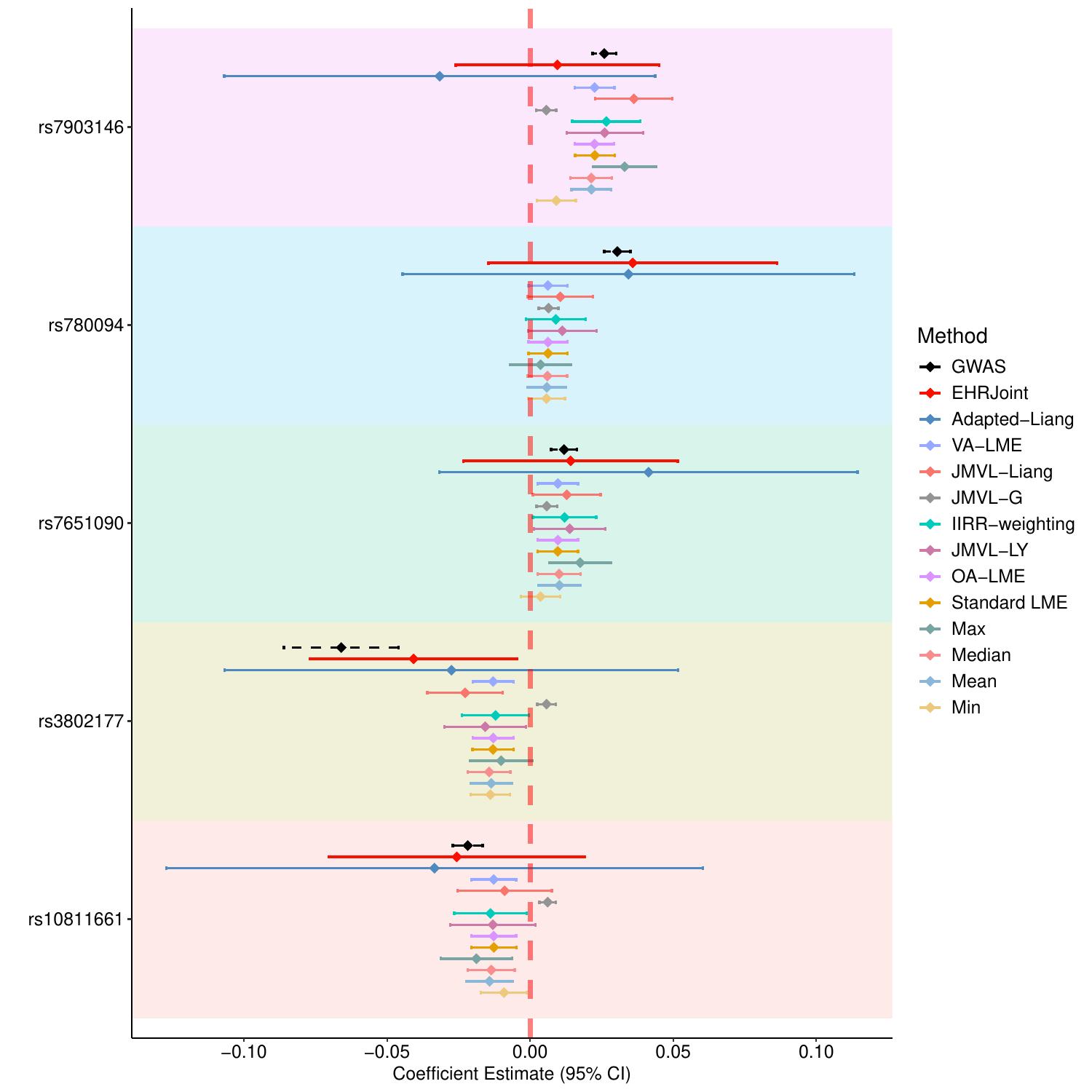}
    \caption{Forest plot of coefficient estimates and 95\% confidence intervals (CIs) for the association between five SNPs and log-transformed glucose levels in the MGI cohort. Values from the GWAS catalog are plotted in dotted lines as reference points.}
    \label{fig:dataExample_results}
\end{figure}

\textbf{Figure \ref{fig:dataExample_results}} presents SNP effect estimates on glucose with 95\% confidence intervals (CIs) across all methods. GWAS benchmark estimates are indicated in dotted lines. The direction of the estimated SNP effects from most methods algins with that of the GWAS, except for Adapted-Liang and JMVL-G. Results from summary statistics methods vary based on the chosen summary metric. Standard LME estimates for rs780094 and rs3802177 differ notably from GWAS, suggesting potential bias arising from IP and IO. Adjusting for the historical number of visits, as in VA-LME, or modeling the visiting process, as in JMVL-Liang, does not fully reduce the bias. In contrast, EHRJoint yields estimates closer to GWAS. 
For example, the estimated effect of rs780094 on log-transformed glucose (mg/dL) is 0.036 (–0.014, 0.086) using EHRJoint, compared to 0.030 (0.026, 0.035) from GWAS. 
In contrast, JMVL-Liang and Standard LME yield estimated effects of 0.010 (-0.001, 0.022) and 0.0062 (-0.0005, 0.0130), respectively. Nonetheless, the impact of IP and IO on point estimates is not always substantial. For rs7903146, rs7651090, and rs10811661, EHRJoint estimates are close to the GWAS benchmarks, suggesting that incorporating IP and IO does not introduce additional bias.

Consistent with simulation results, Standard LME, OA-LME, and VA-LME show smaller variance than the mean summary statistics approach across all SNPs, due to the increased number of observations. JMVL-type methods have higher variance than LME models due to variability from the visiting process, while EHRJoint exhibits the largest variance. For rs780094, EHRJoint yields an SD of 0.026, compared to 0.006 from JMVL-Liang and 0.003 from Standard LME. 

\textbf{(B) Educational disadvantage-glucose association.} 

\begin{figure}
    \centering
    \includegraphics[width=0.7\linewidth]{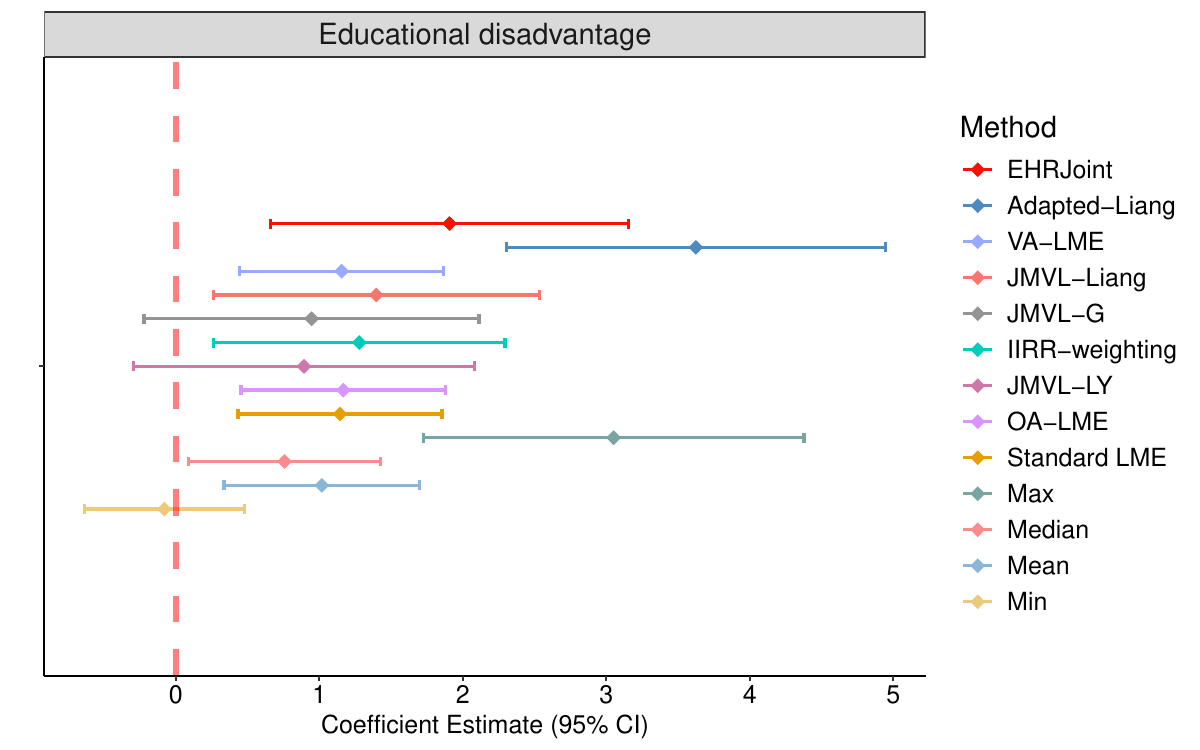}
    \caption{Forest plot of the coefficient estimates and 95\% confidence intervals (CIs) for the association between neighborhood educational disadvantage and glucose levels in the MGI cohort.}
\label{fig:dataExample_results_education}
\end{figure}

In the previous example, the SNP variable is insignificant in the visiting and observation
\newpage 
\noindent 
process models. This motivates us to examine a covariate that is likely to be influencing visiting and observation processes. Educational disadvantage serves as such a variable (\textbf{Figure \ref{fig:dataExample_results_education}}). All methods, except the one using the minimum summary statistics, yield positive estimates for educational disadvantage. This suggests that educational disadvantage is associated with increased glucose levels. Notably, EHRJoint and Adapted-Liang produce larger effect estimates compared to other approaches. 
For example, a one-quartile increase in educational disadvantage is associated with a 1.91 (0.66, 3.16) mg/dL increase in glucose level using EHRJoint and 3.62 (2.31, 4.93) mg/dL using Adapted-Liang. In contrast, JMVL-Liang estimates an increase of 1.40 (0.26, 2.54) mg/dL, while the Standard LME model estimates 1.14 (0.43, 1.85) mg/dL. 
The relatively larger effect estimate from EHRJoint may be due to the significant effect of  educational disadvantage in the observation process, or its correlation with other unmeasured factors influencing patient visits. The variance patterns are similar to those in the SNP analysis, and the findings from the visiting and observation process models remain consistent.

Average runtimes (in seconds) are reported in \textbf{Table S10}. The summary statistics approach, IIRR-weighting, and JMVL-Liang are the fastest, each completing in approximately 2 seconds. EHRJoint takes 16 seconds, which is slightly slower than Standard LME (6s), OA-LME (7s), and VA-LME (6s). In contrast, JMVL-LY (316s) and JMVL-G (1978s) require substantially more time to run.

\section{Conclusion and Recommendations}

Informative presence (IP) and information observation (IO) processes are key challenges in analyzing longitudinal outcomes measured in EHRs. In this paper, we first examine the impact of IP using existing methods in the literature. Our evaluation includes approaches based on summary statistics, mixed-effect models, inverse intensity weighting, and joint modeling of the visiting and longitudinal processes. 
While these methods effectively address the challenge of IP, they are limited when biomarkers are not consistently observed at each 
\newpage 
\noindent visit, i.e., under IO. To simultaneously address IP and IO, we propose a tripartite joint model, EHRJoint, to explicitly model three interdependent processes: the visiting process, the observation process and the longitudinal process. In \textbf{Table \ref{table:summaryOfMethods}}, we provide a summary of the key features and limitations of all methods considered. We develop an R package \texttt{CIMPLE} for \textbf{C}linically \textbf{I}nformative \textbf{M}issingness \textbf{P}rocesses with \textbf{L}ikelihoods and \textbf{E}stimating equations to implement all methods.

\noindent \textbf{\underline{When to model the visiting process}}

Our results demonstrate that modeling the visiting process is unnecessary when IP is absent or when the longitudinal and visiting processes are related only through measured variables. In these cases, simpler approaches (e.g., mean summary statistics or mixed-effects models) provide unbiased estimates of the exposure effects while avoiding the computational cost of joint modeling. This aligns with prior work showing that conditional independence between the visiting and longitudinal processes eliminates bias in coefficient estimates \citep{neuhaus2018analysis,gasparini2020mixed}. 
However, this assumption may not hold in real-world EHR data, where patient visits are often influenced by unmeasured or latent factors through complicated mechanisms \citep{hatef2019assessing,lasser2021social}. In such settings, failure to account for IP can introduce substantial bias, and joint modeling becomes necessary. Among existing methods, JMVL-Liang yields the smallest bias due to its inclusion of dependent latent variables. 

\noindent \textbf{\underline{The importance of addressing IO}}

Our study highlights the need to also consider the observation process of the biomarker.
In practice, not all clinic visits result in biomarker measurements, and the reasons for missing data are often multifaceted and not random \citep{groenwold2020informative,li2021imputation}. Existing joint modeling methods, such as JMVL-Liang, treat all patients without biomarker measurements as ``missing'', implicitly assuming a homogeneous missingness mechanism. However, this approach overlooks key heterogeneity between patients who have clinic visits but no biomarker recorded, and those with no visits at all. The former group may consist of individuals with a higher disease burden requiring frequent visits, while the latter may include relatively healthier individuals \citep{rusanov2014hidden,palladino2016associations}. 

As a result, when both IP and IO are present, existing methods produce biased coefficient estimates because they ignore the observation process. Our proposed method, EHRJoint, is the only method that remains unbiased in the presence of IP and IO. Notably, even when the visiting process is misspecified, our proposed method yields the smallest bias across all methods considered. However, explicitly modeling all the three processes comes with a trade-off: while reducing bias, modeling the visiting and observation processes increases the variance of coefficient estimators to ensure proper coverage of confidence intervals.

\noindent \textbf{\underline{Limitations and extensions}}

The consistency of the coefficient estimators in the longitudinal model $\boldsymbol{\beta}$ in EHRJoint relies on the validity of the estimating equation \eqref{Estimating_eq_betatheta}, which is derived under the specified three models. Notably, correct specification of the visiting process and the longitudinal outcome model is crucial to ensuring valid estimation of $\boldsymbol{\beta}$. In contrast, the specification of the biomarker observation process allows for more flexibility as long as the observation probability $\omega_i(t)$ is consistently estimated. While we do not provide theoretical guarantees on the robustness of our method, we conducted simulation studies to empirically assess its 
\newpage 
\noindent performance: Even when the visiting process model was misspecified, EHRJoint consistently demonstrated the smallest bias across all misspecified scenarios.

Furthermore, in this paper, we implement a logistic regression model with time-invariant covariates for the observation process of the biomarker. While it provides a fundamental framework of understanding IO, incorporating time-varying covariates could provide a deeper understanding of the dynamic factors influencing biomarker observation \citep{tan2023informative}. For example, a patient's evolving clinical status (e.g., disease progression, medication use, or recent hospitalizations) may affect the likelihood of biomarker measurement. Extending this framework to accommodate time-varying covariates through generalized mixed-effects models—with potential integration of shared random effects across all three sub-models—would be an exciting direction for future research.

\begin{table}[H]
\scriptsize
\caption{Summary of methods for analyzing longitudinal measurements in the EHR with a focus on estimating fixed effect parameters corresponding to measured exposures and covariates in the longitudinal model. IP: informative presence. IO: informative observation.}
\label{table:summaryOfMethods}
\begin{tabularx} {\columnwidth}{yxzyx} \Xhline{2\arrayrulewidth} \hline
   \textbf{Method}    &  \textbf{IP/IO} & \textbf{Features} & \textbf{Limitations} & \textbf{Package}  \\ \Xhline{2\arrayrulewidth}
\multirow{2}{*}{\begin{tabular}[l]{@{}l@{}} Standard LME \\ \cite{laird1982random} \end{tabular}}  & neither &
Provides consistent coefficient estimates in the absence of IP and IO, or when the visiting process depends on the longitudinal process through measured variables. More efficient than using the mean summary statistics.&
Produces biased coefficient estimates of covariates when IP and/or IO arises from unmeasured factors.  &
\href{https://cran.r-project.org/web/packages/lme4/index.html}{lme4}. \\ \hline
\multirow{2}{*}{\begin{tabular}[l]{@{}l@{}}  Visit-adjusted LME \\ \cite{goldstein2016controllingEHRencounter} \end{tabular}}    & neither &
Adjusts for the time-varying number of visits in the longitudinal model. &
Simulation studies show no improvement over the Standard LME. &
\href{https://cran.r-project.org/web/packages/lme4/index.html}{lme4}. \\ \hline
\multirow{3}{*}{\begin{tabular}[l]{@{}l@{}l@{}} Observation-adjusted \\ LME  \\ \cite{neuhaus2018analysis}  \end{tabular}} 
& neither &
Adjusts for the time-varying number of observations in the longitudinal model. &
Simulation studies show no improvement over the Standard LME. &
\href{https://cran.r-project.org/web/packages/lme4/index.html}{lme4}. \\ \hline
\multirow{2}{*}{\begin{tabular}[l]{@{}l@{}} JMVL-LY  \\ \cite{lin2001semiparametric}\end{tabular}} 
 & IP &
Models the visiting process by a counting process, with dependence on the longitudinal process through shared measured variables. &
Increased bias and variance compared to the Standard LME. Computationally intense. & 
- \\ \hline
\multirow{2}{*}{\begin{tabular}[l]{@{}l@{}} JMVL-Liang \\ \cite{liang2009joint}  \end{tabular}} 
& IP &
Links the visiting and longitudinal processes through dependent latent variables. Simulation study shows the smallest bias among all methods across all IP-only simulation scenarios. & Exhibits larger variance than mixed-effect models due to the added variability from the visiting process. & 
\href{https://cran.r-project.org/web/packages/IrregLong/index.html}{IrregLong}. \\ \hline
\multirow{2}{*}{\begin{tabular}[l]{@{}l@{}} JMVL-G  \\ \cite{gasparini2020mixed} \end{tabular}} 
& IP &
Models the visiting process using a proportional rate model for the inter-visit time. The longitudinal and visiting processes share a random intercept. & Only corrects bias in estimating the intercept under IP, with little impact on other coefficient estimates. Computationally intense. & \href{https://cran.r-project.org/web/packages/merlin/index.html}{merlin}.
\\ \hline
\multirow{2}{*}{\begin{tabular}[l]{@{}l@{}} IIRR-weighting  \\ \cite{burvzkova2007longitudinal} \end{tabular}}
 & IP &
Addresses IP through weighting, where observations are weighted based on the inverse of their recording intensity. & Simulation studies show a similar performance to JMVL-LY.
& 
\href{https://cran.r-project.org/web/packages/IrregLong/index.html}{IrregLong}. \\ \hline
EHRJoint 
& IP and IO & The only unbiased method under both IP and IO. Computationally efficient.
& Increased variance due to the added variability from the visiting and observation process.
& -
 \\ \hline
\Xhline{2\arrayrulewidth}    
\end{tabularx}
\end{table}

\bibliographystyle{apalike}  
\bibliography{references}  

\section*{Acknowledgment}

We thank Howard Baik, Data Scientist at Yale University, Public Health Data Science and Data Equity, for their help with R coding and the development of the R package.

\end{document}